\documentclass[%
 reprint,
 amsmath,amssymb,
 aps,
floatfix,
pre
]{revtex4-2}

\usepackage{graphicx}% Include figure files
\usepackage{dcolumn}% Align table columns on decimal point
\usepackage{bm}% bold math
\graphicspath{ {./images/} }
\usepackage{wrapfig}
\usepackage{enumitem}
\usepackage{amsmath,stackengine}
\stackMath
\usepackage{amssymb}
\usepackage{marginnote}
\usepackage{comment}
\usepackage{url}
\usepackage{esint}
\usepackage{multirow}
\usepackage[caption=false]{subfig}
\usepackage[dvipsnames]{xcolor}
\usepackage[colorlinks, allcolors=NavyBlue]{hyperref}
\usepackage{tikz}
\usetikzlibrary{arrows.meta}

\usepackage{mathtools}
\usepackage{listings}
\usepackage{pgfplots}
\pgfplotsset{compat=1.9, , every axis/.append style={font=\scriptsize}}
\usepackage{xspace}
\usepackage{mathrsfs}
\usepackage{physics}

\begin{document}

\title{Phase-space networks and connectivity of the kagome antiferromagnet}

\author{Brandon B. Le}
\author{Seung-Hun Lee}
\author{Gia-Wei Chern}
\affiliation{
Department of Physics, University of Virginia, Charlottesville, Virginia 22904, USA
}

\date{\today}% It is always \today, today,
             %  but any date may be explicitly specified

\begin{abstract}
We study the coplanar ground-state manifold of the kagome Heisenberg antiferromagnet using a phase-space network representation, in which nodes correspond to coplanar ground states and edges represent transitions generated by weathervane loop rotations. In the coplanar manifold, each configuration can be mapped to a three-coloring problem on the dual honeycomb lattice, where a weathervane mode corresponds to a closed loop of two alternating colors. By comparing networks that include all weathervane loops with networks restricted to elementary six-spin loops, we examine how energetic constraints shape phase-space structure. We find that connectivity distributions are sharply peaked in large systems, while restrictions to short loops reduce typical connectivity. Spectral properties further distinguish the two cases, with short-loop networks exhibiting Gaussian spectra and full networks displaying non-Gaussian features associated with correlated loop updates. Finally, a box-counting analysis reveals distinct finite-size geometric trends: short-loop networks show algebraic, fractal-like behavior, whereas full networks are more compact in graph distance due to longer-loop shortcuts. These results show that the hierarchy of weathervane loop rotations provides a direct link between microscopic constraints and emergent phase-space geometry in a frustrated magnet.
\end{abstract}

%\keywords{Suggested keywords}%Use showkeys class option if keyword
                              %display desired
\maketitle

%\tableofcontents

\newpage

\section{Introduction}

Geometrical frustration arises when lattice geometry prevents all local interaction constraints from being satisfied simultaneously, leading to an extensive degeneracy of low-energy states~\cite{wannier1950antiferromagnetism,Anderson56,pauling1935structure}. This macroscopic degeneracy suppresses conventional long-range order and gives rise to unconventional thermodynamic behavior, slow relaxation dynamics, and emergent collective phenomena. Prominent realizations occur in magnetic materials, ranging from naturally occurring compounds to engineered artificial spin systems~\cite{Lacroix2011,bramwell2001spin,Nisoli13,Udagawa2021}. In such systems, the absence of conventional ordering often stabilizes cooperative paramagnets or spin liquids, characterized by strong short-range correlations persisting to the lowest temperatures~\cite{Ramirez1994,Moessner98,Henley2010CoulombPhase,Balents2010}. Owing to their strong local constraints, frustrated magnets are also closely connected to broader classes of constrained models---such as vertex, dimer, and ice models---that provide minimal frameworks for understanding frustration-induced collective behavior~\cite{lieb1967exact,lieb1967residual,eloranta1999diamond}.

Among frustrated magnets, the kagome Heisenberg antiferromagnet occupies a central place in the development of the modern theory of geometrical frustration~\cite{Huse1992,zeng1990numerical,chalker1992hidden,harris1992possible}. Composed of corner-sharing triangles, the kagome lattice maximally frustrates antiferromagnetic interactions, resulting in an extensive classical ground-state degeneracy and a strong suppression of long-range order. Early experimental studies of SCGO, SrCr$_{9p}$Ga$_{12-9p}$O$_{19}$, which realizes a kagome bilayer rather than an ideal single kagome plane, were instrumental in establishing the field by revealing persistent spin fluctuations and cooperative paramagnetism down to very low temperatures~\cite{ramirez1990strong,uemura1994spin}. More direct single-layer kagome realizations include the $S=1/2$ quantum kagome antiferromagnet herbertsmithite, $\mathrm{ZnCu}_3(\mathrm{OH})_6\mathrm{Cl}_2$~\cite{shores2005structurally,helton2007spin,han2012fractionalized}, as well as larger-spin kagome compounds such as the jarosites, e.g. $\mathrm{KFe}_3(\mathrm{OH})_6(\mathrm{SO}_4)_2$~\cite{matan2006spin}. Beyond its historical importance, the kagome antiferromagnet has become a paradigmatic platform for order-by-disorder phenomena: both thermal and quantum fluctuations select coplanar spin configurations from the degenerate classical manifold, as these states maximize entropy and minimize the zero-point energy of magnons~\cite{harris1992possible,chalker1992hidden,Reimers1993,Zhitomirsky2008,taillefumier2014semiclassical,chubukov1992order,chubukov1993order}.

Despite this confinement to coplanar configurations, the coplanar classical ground-state manifold itself remains highly degenerate. Advanced Monte Carlo simulations have shown that thermal fluctuations beyond the harmonic approximation further lift this residual degeneracy and ultimately select the $\sqrt{3}\times \sqrt{3}$ magnetic order~\cite{Henley2009,Chern2013}. Quantum fluctuations beyond linear spin-wave theory similarly favor the same ordering pattern, as demonstrated by nonlinear spin-wave calculations and exact diagonalization studies~\cite{chubukov1992order,chubukov1993order,Sachdev1992,Cepas2011,Gotze2011,chernyshev2014quantum}. In real materials, however, additional perturbations and dynamical effects—such as disorder, further-neighbor interactions, or coupling to other degrees of freedom—often dominate over these higher-order order-by-disorder mechanisms in determining the ultimate ordering pattern.

Nonetheless, extensive prior work has firmly established the coplanar manifold as the essential low-energy arena in which novel emergent ordering tendencies and dynamical phenomena, including glassy behavior and heterogeneous dynamics, can arise~\cite{cepas2012heterogeneous,cepas2014multiple}. This perspective is further reinforced by the fact that the coplanar manifold itself constitutes a classical critical state with algebraic spin–spin correlations. Indeed, the coplanar ground states are exactly equivalent to the three-coloring problem on the honeycomb lattice, dual to the kagome lattice~\cite{Baxter1970,Kondev96,Chakraborty2002,Castelnovo2005,Cepas2017}. This canonical model provided one of the earliest exactly or quasi-exactly tractable examples demonstrating that criticality can emerge purely from local constraints rather than fine-tuned interactions, helping to establish the modern notion of constraint-induced critical phases.

\begin{figure*}[htbp!]
    \subfloat[]{\includegraphics[width=0.33\textwidth]{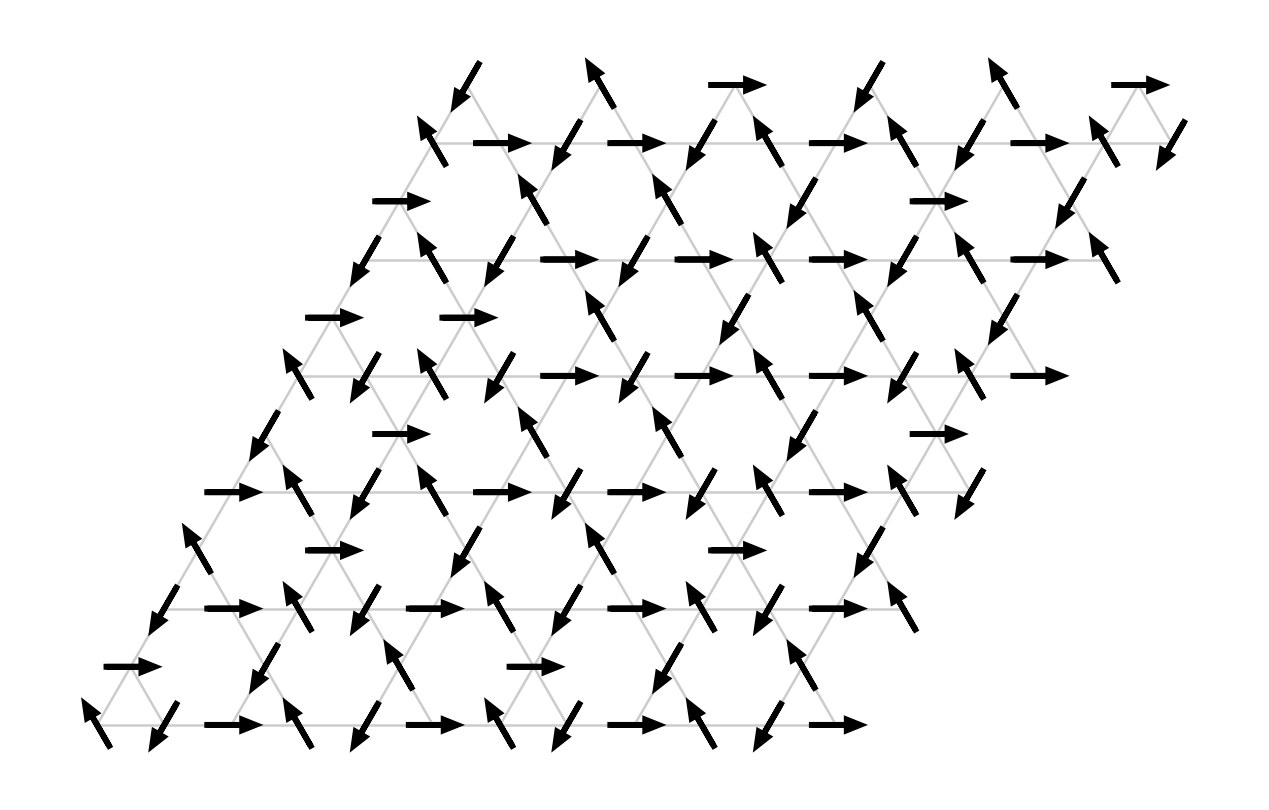}\label{fig:schematic_arrows}}
    \hfill
    \subfloat[]{\includegraphics[width=0.33\textwidth]{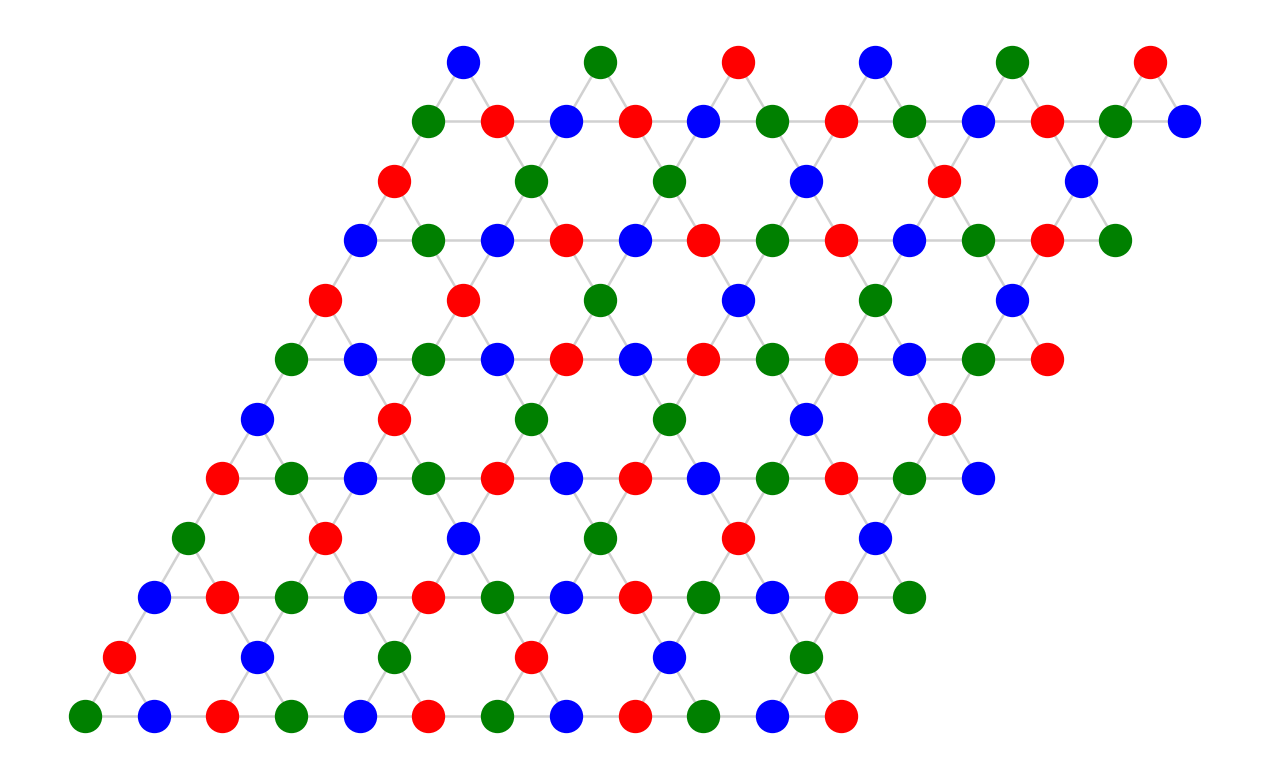}\label{fig:schematic_potts}}
    \hfill
    \subfloat[]{\includegraphics[width=0.33\textwidth]{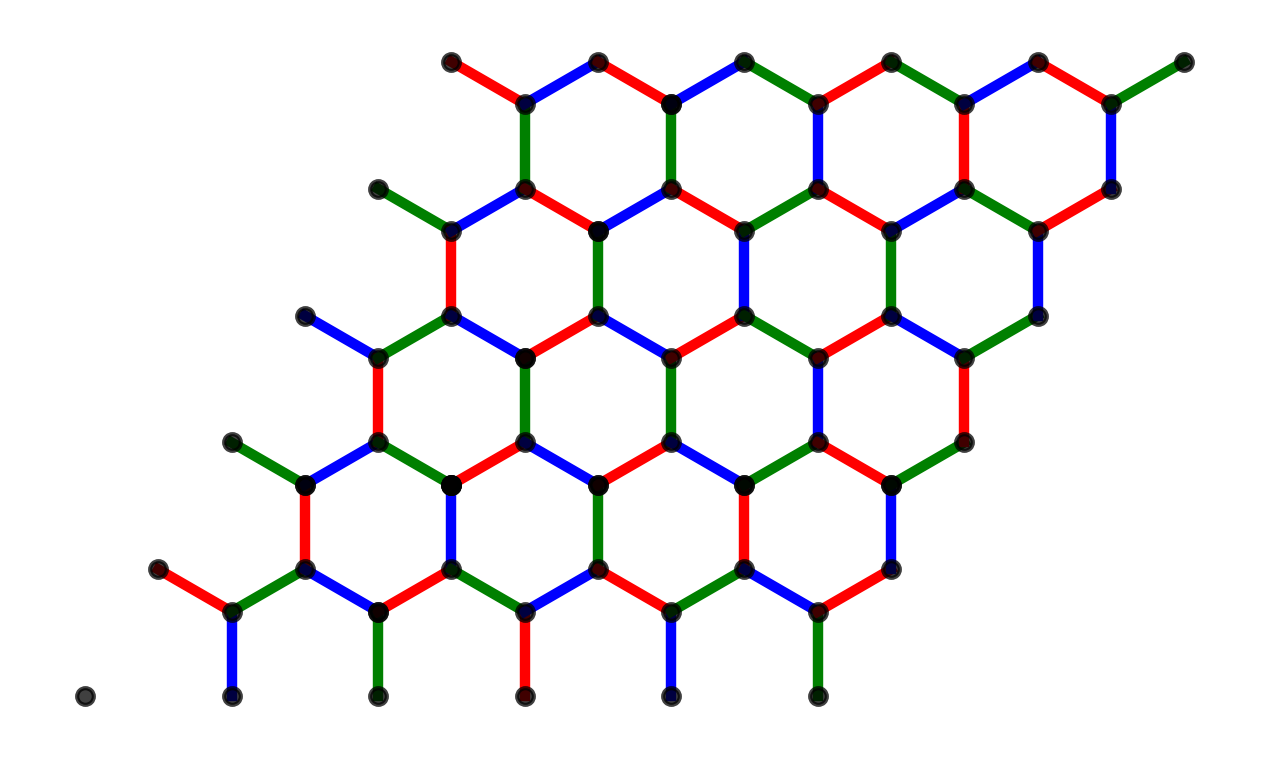}\label{fig:schematic_honeycomb}}
    \caption{Three equivalent representations of a random coplanar ground state of a $6\times 6$ kagome lattice, i.e., $6\times6$ unit cells or $N=108$ kagome spins, with periodic boundary conditions. (a)~A coplanar spin configuration on the kagome lattice, with arrows indicating spin directions separated by $120^\circ$. (b)~The same state represented as an antiferromagnetic three-state Potts configuration on the kagome lattice, where each site is assigned one of three labels $\{R, G, B\}$ corresponding to the coplanar spin orientations. (c)~The associated three-coloring problem on the dual honeycomb lattice: each kagome site maps to an edge of the honeycomb lattice, and the Potts label of that site determines the color of the corresponding dual edge.}
    \label{fig:schematic}
\end{figure*}

The three-coloring problem on the honeycomb lattice admits a continuum description in terms of a discrete height model defined on the dual triangular lattice, where the local coloring constraint enforces a divergence-free condition on the height field~\cite{Kondev96,Chakraborty2002,Chern2013}. This constraint is analogous to the local Gauss-law constraints that arise in emergent gauge descriptions of Coulomb-phase systems~\cite{Henley2010CoulombPhase}. Upon coarse-graining, this mapping leads to a critical Gaussian free-field theory with a symmetry-fixed stiffness, placing the system in a Coulomb phase without fine tuning. As a result, the model exhibits algebraic correlations, with color and bond observables decaying as power laws governed by the scaling dimensions of operators in the height theory.

In this work, we analyze the coplanar ground-state manifold of the kagome antiferromagnet---equivalently, the configuration space of the three-coloring problem---using the framework of phase-space networks. This network-based perspective provides a natural and complementary viewpoint to field-theoretical approaches for characterizing this highly degenerate and constrained system. Phase-space network representations have proven powerful in frustrated and constrained systems, where configuration spaces are large, discrete, and structured by local rules~\cite{han2009phase,han2010phase,peng2011self,lee2021frustration,cao2015ground}. By mapping phase space onto a graph, network-theoretic tools can be used to characterize connectivity, heterogeneity, spectral properties, and scaling behavior within a unified framework~\cite{strogatz2001exploring,albert2002statistical,dorogovtsev2002evolution,newman2003structure,boccaletti2006complex,st2022analysis}. Such approaches have revealed universal statistical features, including Gaussian connectivity distributions~\cite{costa2007characterization,javarone2015gaussian,chang2025degree}, characteristic spectral densities~\cite{farkas2001spectra,de2005spectral,metz2020spectral}, and nontrivial geometric properties such as fractal structure and self-similarity~\cite{hutchinson1981fractals,theiler1990estimating,gallos2007review,fronczak2024scaling,bunimovich2024fractal}.

In the kagome antiferromagnet, coplanar ground states are connected by collective weathervane-loop rotations spanning a hierarchy of length scales and associated energy barriers~\cite{cepas2012heterogeneous,cepas2014multiple,le2026energy}. A phase-space network therefore provides a natural framework for encoding both local mobility and the global organization of the coplanar manifold. To isolate the role of barrier-energy hierarchies in shaping phase-space structure and dynamics, we analyze both the full phase-space network---including weathervane loops of all allowed lengths---and a restricted network containing only elementary six-spin loops. We characterize these networks through their connectivity distributions, adjacency spectra, and fractal box-counting properties. By systematically comparing the full and short-loop networks across system sizes, we demonstrate how long-loop updates qualitatively reshape phase-space connectivity and enhance the effective dimensionality of configuration space.

The remainder of this paper is organized as follows. In Sec.~\ref{sec:networks}, we construct the phase-space networks of coplanar ground states. In Sec.~\ref{sec:connectivity}, we analyze the networks' connectivity distributions and their scaling behavior with system size. Sec.~\ref{sec:spectral} examines the spectral properties of the networks and relates them to closed-walk statistics and collective connectivity modes. In Sec.~\ref{sec:fractal}, we characterize the geometric structure of phase space using fractal box-counting analysis. Finally, Sec.~\ref{sec:conclusions} summarizes our results and discusses their broader implications.

\begin{figure*}[htbp!]
    \subfloat[]{\includegraphics[width=0.33\textwidth]{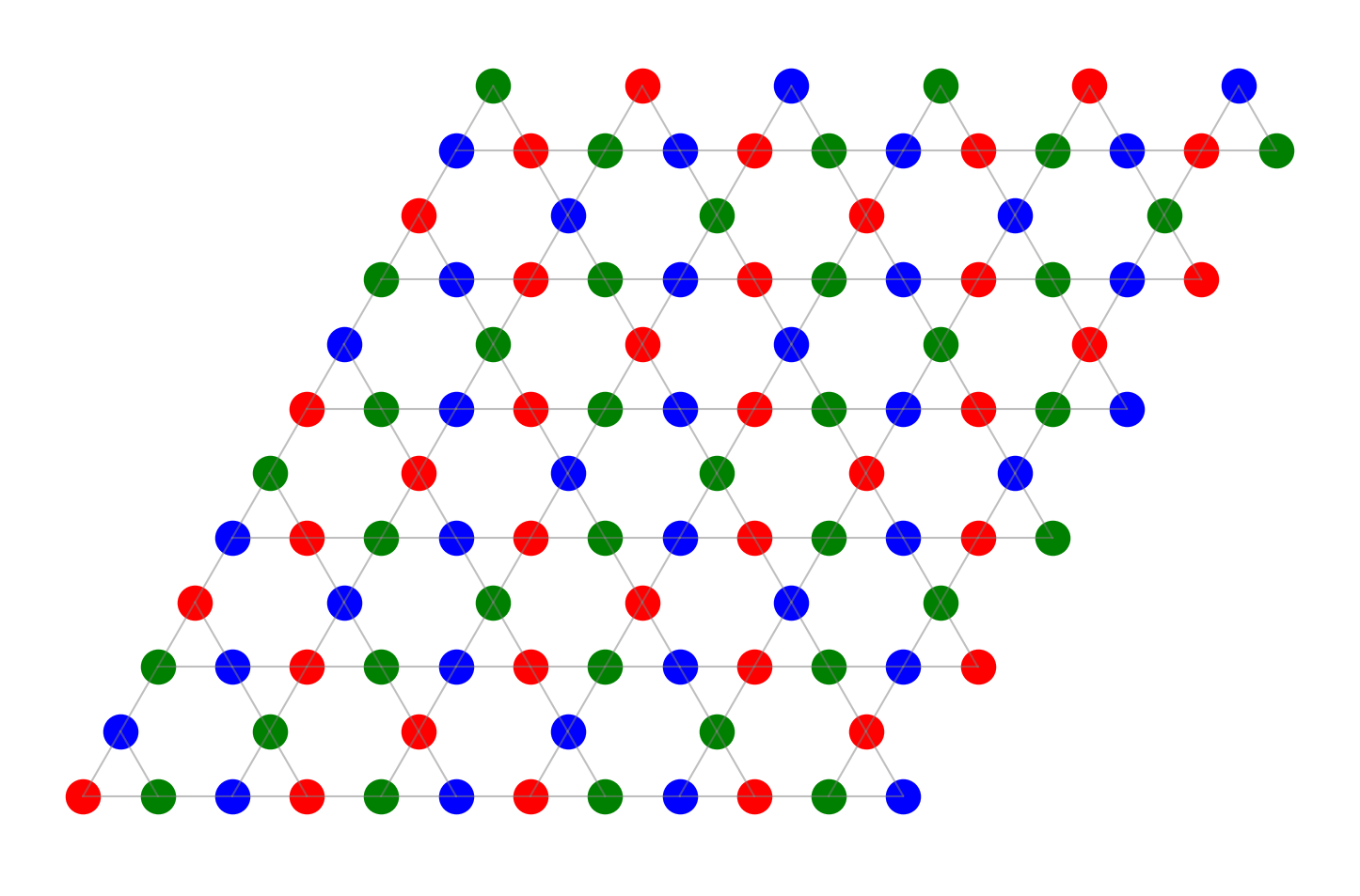}\label{fig:lattices-a}}
    \hfill
    \subfloat[]{\includegraphics[width=0.33\textwidth]{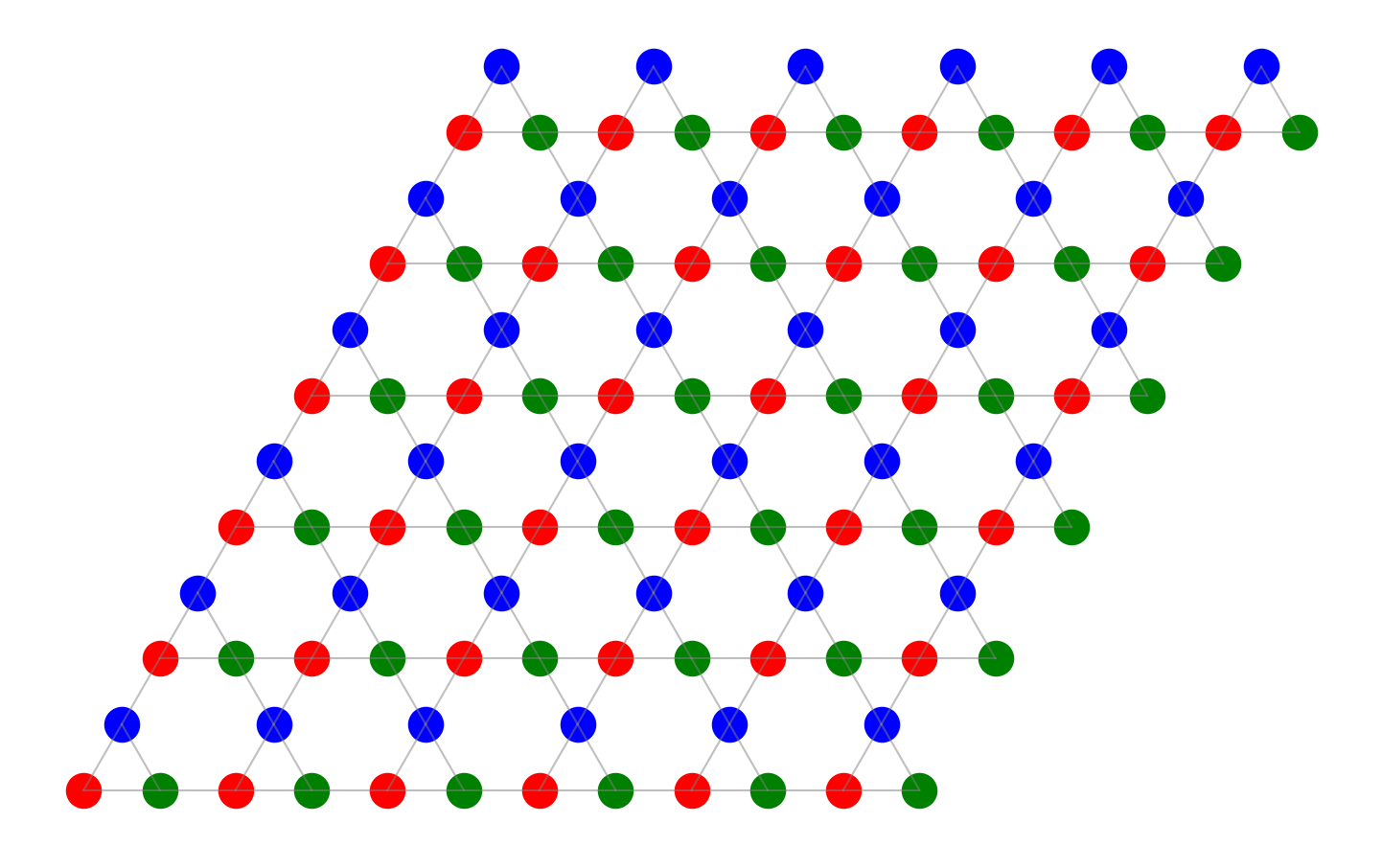}\label{fig:lattices-b}}
    \hfill
    \subfloat[]{\includegraphics[width=0.33\textwidth]{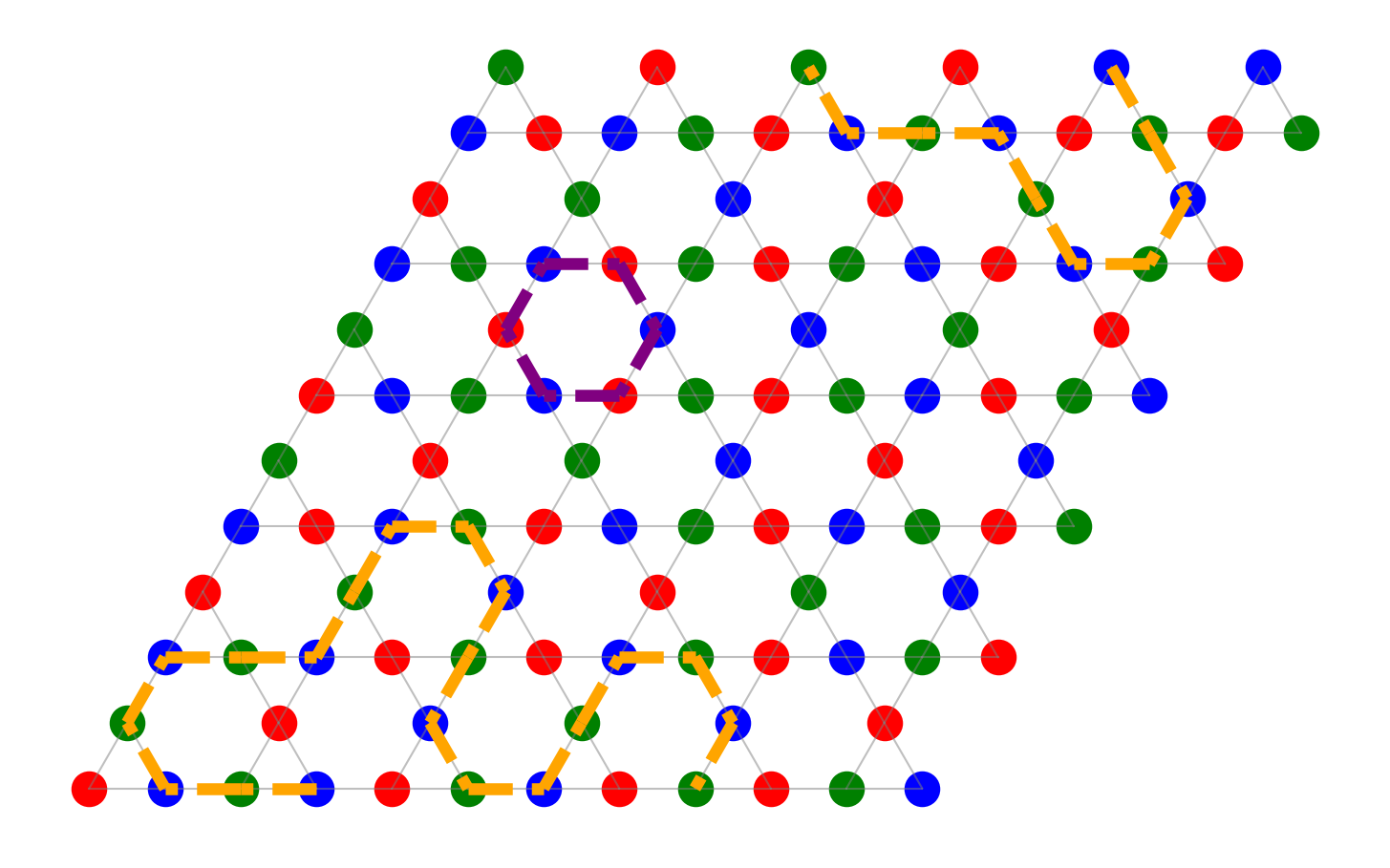}\label{fig:lattices-c}}
    \caption{Examples of coplanar ground states of a $6\times 6$ unit-cell kagome lattice with periodic boundary conditions, shown in the three-state Potts representation where each color corresponds to one of the three spin orientations separated by $120^\circ$. (a)~A $\sqrt{3}\times\sqrt{3}$ state. (b)~A $q=0$ state. (c)~A representative disordered coplanar configuration, with an elementary six-spin weathervane loop (purple) and a longer weathervane loop (orange) highlighted.}
    \label{fig:lattices}
\end{figure*}

\section{Kagome Phase-Space Networks}
\label{sec:networks}

We consider the classical nearest-neighbor Heisenberg antiferromagnet
\begin{eqnarray}
	\mathcal{H} = J \sum_{\langle ij \rangle} \mathbf S_i \cdot \mathbf S_j 
\end{eqnarray}
on finite $L_x\times L_y$ kagome lattices with periodic boundary conditions. Throughout this work, $L_x$ and $L_y$ denote the numbers of kagome Bravais unit cells along the two primitive directions. Since each kagome unit cell contains three sites, the total number of spins is $N=3L_xL_y$. The kagome lattice possesses an extensively degenerate ground-state manifold arising from geometric frustration~\cite{Huse1992,zeng1990numerical,chalker1992hidden,harris1992possible}. Any spin configuration satisfying the local constraint that the vector sum of spins vanishes on every triangle is a classical ground state, leading to an exponential growth of degeneracy with system size. Both thermal and quantum fluctuations partially lift this degeneracy and robustly select coplanar spin configurations from the continuous classical manifold~\cite{harris1992possible,chalker1992hidden,Reimers1993,Zhitomirsky2008,taillefumier2014semiclassical,chubukov1992order,chubukov1993order}. In the coplanar manifold of the kagome Heisenberg antiferromagnet, all spins lie within a common plane and, on every elementary triangle, are arranged at mutual angles of $120^\circ$; see Fig.~\ref{fig:schematic_arrows} for an example. Such configurations exactly satisfy the classical ground-state constraint that the vector sum of spins vanishes on each triangle. While coplanarity is enforced, the manifold remains extensively degenerate because the specific assignment of the three $120^\circ$-separated spin directions can vary from triangle to triangle, subject only to local consistency constraints imposed by shared spins.

This structure admits a natural mapping to the antiferromagnetic three-state Potts model. Each spin is assigned one of three discrete states, conveniently labeled by the colors $R$, $G$, and $B$, corresponding to the three allowed spin orientations in the coplanar plane; see Fig.~\ref{fig:schematic_potts}. The kagome constraint that the three spins on every triangle must all differ translates directly into the Potts constraint that all three colors appear exactly once on each triangle. As a result, the coplanar manifold of the kagome antiferromagnet is isomorphic to the ground-state manifold of the antiferromagnetic three-state Potts model.

Equivalently, this Potts description can be reformulated as a three-coloring problem on the dual honeycomb lattice, a well-studied highly constrained system~\cite{Baxter1970,Kondev96,Chakraborty2002,Castelnovo2005,Cepas2017}. In this representation, each kagome spin corresponds to a colored bond of the honeycomb lattice, and the local kagome constraint enforces that the three bonds meeting at every honeycomb vertex carry distinct colors, as illustrated in Fig.~\ref{fig:schematic_honeycomb}. This mapping provides a purely discrete, constraint-based description of the coplanar manifold and underlies its further characterization in terms of height models, critical correlations, and phase-space structure~\cite{Kondev96,Chakraborty2002,Chern2013}.

Within the coplanar ground-state manifold, both highly ordered and disordered configurations occur. Two well-known periodic states are the $q=0$ state, in which the three colors repeat uniformly across the lattice, and the $\sqrt{3}\times\sqrt{3}$ state, which enlarges the magnetic unit cell and maximizes the number of flippable loops~\cite{chubukov1992order,chubukov1993order}. In contrast, generic coplanar states display irregular color patterns without long-range periodicity. Figure~\ref{fig:lattices} presents representative examples of these coplanar configurations in the three-state Potts representation. Although they are energetically degenerate at harmonic order, these states differ markedly in their local color arrangements, which in turn control the allowed collective rearrangements.

Distinct coplanar configurations are connected by collective local modes known as weathervane loop rotations~\cite{henley2001effective,landau2012spin,taillefumier2014semiclassical}. A weathervane loop is a closed path of alternating colors (Fig.~\ref{fig:lattices-c}); spins along the loop can be continuously rotated about the axis defined by the third, absent color without violating local constraints. Such rotations map one coplanar ground state to another and constitute the elementary moves that generate connectivity within the coplanar manifold. In the language of the three-coloring problem, a weathervane loop corresponds to a closed loop of links with two alternating colors on the honeycomb lattice. From this perspective, the coplanar ground-state manifold can naturally be viewed as a network in phase space: each node represents a distinct coplanar ground state, and an edge connects two nodes if they are related by a single weathervane loop rotation.

Each edge in the phase-space networks is assigned an energy associated with the weathervane-loop rotation connecting the two coplanar configurations. To compute this quantity, we parametrize the continuous loop rotation by an angle $\theta$ and evaluate the harmonic zero-point energy along the rotation path. The spins are mapped to bosonic operators via the Holstein--Primakoff transformation, and the resulting quadratic bosonic Hamiltonian is diagonalized using a real-space Bogoliubov transformation. This gives a zero-point energy per spin $E(\theta)$ for each intermediate configuration. We define the barrier energy of the loop as
\begin{equation}
    E_{\rm b} = \max_{\theta} E(\theta),
\end{equation}
namely, the highest energy per spin encountered during the weathervane-loop rotation.

It is useful to distinguish the equilibrium and dynamical roles of this network representation. The nodes represent physical coplanar configurations. Within the classical nearest-neighbor model, all such states have the same classical energy, and they remain degenerate at harmonic order in the spin-wave expansion. Thus, at the level considered here, a restricted partition function over the coplanar manifold takes the form
\begin{equation}
    Z_{\rm cop}^{(2)} = \sum_{\mathcal C}e^{-\beta\left(E_{\rm cl}+E_{\rm zp}^{(2)}\right)} = \mathcal N_{\rm cop}e^{-\beta\left(E_{\rm cl}+E_{\rm zp}^{(2)}\right)},
\end{equation}
where $\mathcal C$ labels coplanar configurations, $\mathcal N_{\rm cop}$ is the number of such configurations, and $E_{\rm zp}^{(2)}$ is the common harmonic zero-point energy of the coplanar states. The nontrivial energy scale in our network is instead associated with the edges: $E_{\rm b}$ measures the highest harmonic zero-point energy encountered along the continuous weathervane-loop rotation path connecting two degenerate coplanar endpoints. These edge weights therefore characterize the relative difficulty of moving between neighboring coplanar configurations, rather than defining different equilibrium statistical weights for the nodes. In principle, they could be used to define a finite-temperature Monte Carlo or kinetic Monte Carlo dynamics on the full weathervane-loop network, in which higher-barrier loops become progressively less accessible as temperature is lowered. Consequently, the temperature dependence discussed below should be understood as a dynamical restriction associated with time-scale separation, not as a thermodynamic phase transition. In particular, this interpretation does not conflict with the Mermin--Wagner theorem, since no finite-temperature long-range magnetic order or spontaneous breaking of the continuous spin-rotation symmetry is being claimed~\cite{mermin1966absence}.

With this distinction in mind, the low-temperature dynamics within this manifold (equivalently, within the space of three-coloring configurations) is strongly constrained by the structure of the allowed collective moves. Transitions between coplanar states occur exclusively through weathervane loop rotations, which are thermally activated processes. The associated energy barrier increases with the loop length, leading to transition rates that are exponentially suppressed for long loops. Consequently, the dynamics develops a pronounced hierarchy of time scales, with short loops governing fast relaxation processes and longer loops controlling increasingly rare and slow rearrangements~\cite{cepas2012heterogeneous,cepas2014multiple}.

At sufficiently low temperatures, this hierarchy has a profound impact on phase-space exploration~\cite{cepas2012heterogeneous}. Short weathervane loops—most notably the elementary six-spin loops—remain dynamically active and efficiently thermalize subsets of coplanar states. In contrast, transitions requiring longer loops become effectively frozen on accessible time scales. The phase space therefore fragments into trapping sectors: within each sector, states are rapidly equilibrated by short-loop dynamics, while equilibration between different sectors requires collective rearrangements involving longer loops and is suppressed at low temperature. This separation of time scales gives rise to slow relaxation, dynamical heterogeneity, and effective ergodicity breaking on experimental or simulation time scales.

This constrained dynamics motivates two complementary network representations of the coplanar manifold. In the full phase-space network, each node corresponds to a coplanar ground state, and edges represent all allowed weathervane loop rotations, weighted by their associated energy barriers (or equivalently, inverse transition rates). This weighted network encodes the complete hierarchy of dynamical processes, ranging from fast short-loop updates to rare long-loop transitions that connect distant regions of phase space.

At low temperatures, however, the dynamics is effectively restricted to a reduced subnetwork. A short-loop network can be defined by retaining only edges associated with the shortest weathervane loops, namely the length-six loops. This network generally decomposes into multiple disconnected components, each corresponding to a trapping sector that can internally thermalize but remains dynamically isolated from others on relevant time scales~\cite{cepas2012heterogeneous}. The appearance of these disconnected components provides a direct, graph-theoretic representation of trapping and slow dynamics, and establishes a transparent link between constrained microscopic motion and the fragmented structure of phase space emphasized in this work.

A closely related viewpoint is that dynamics dominated by the shortest loops can leave geometrically “insulated” regions trapped in configuration space, while the activation of longer loops progressively reconnects these regions and introduces slower relaxation channels~\cite{cepas2014multiple}. Within our network framework, the short-loop network thus serves as a structural proxy for the fast, locally mobile sector of the dynamics, whereas the full network captures how longer-loop processes restore global connectivity and govern long-time relaxation.

\begin{figure*}[htp!]
    \subfloat[]{\includegraphics[width=0.55\textwidth]{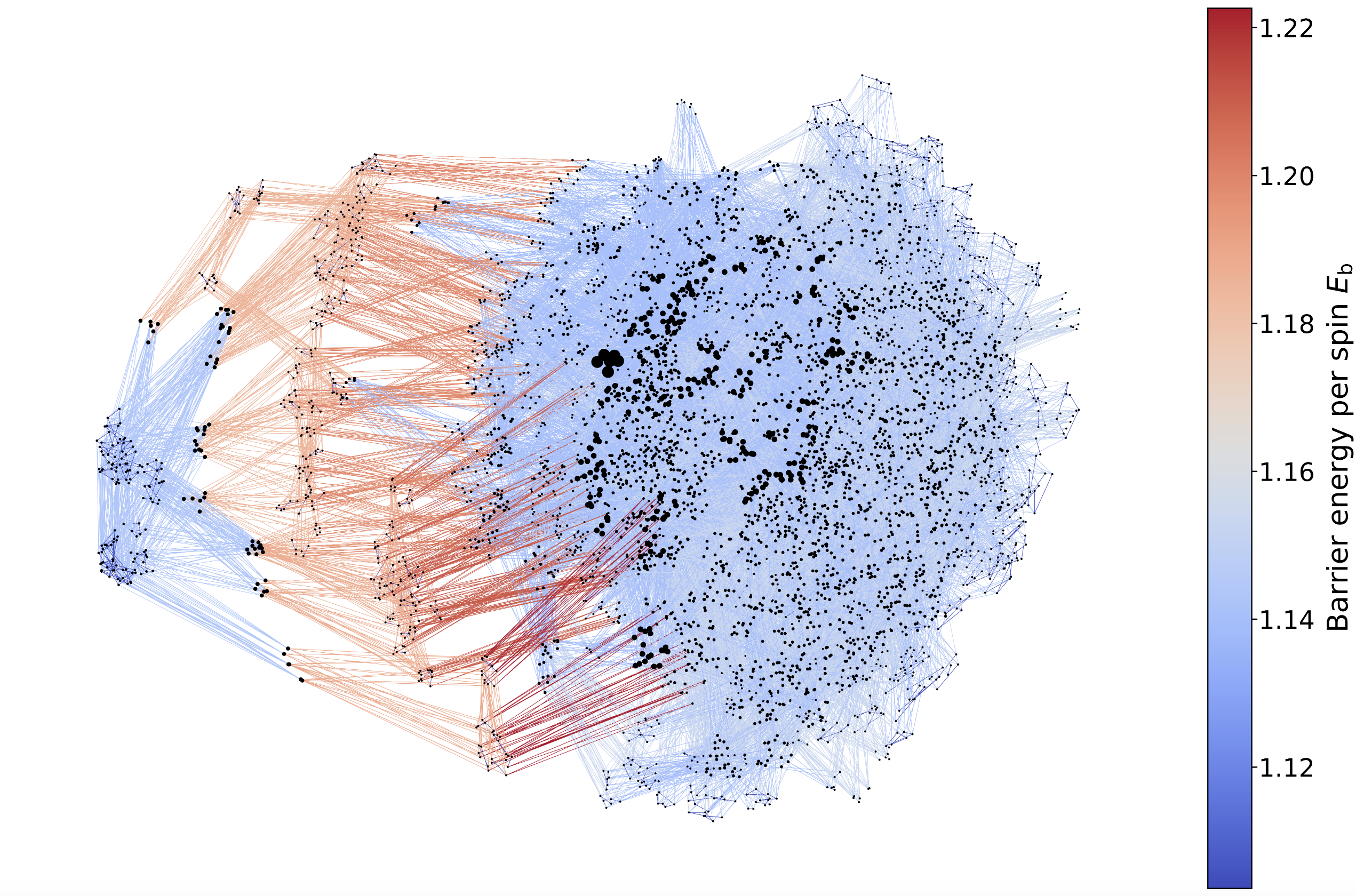}\label{fig:6x3_graphs-a}}
    \hfill
    \subfloat[]{\includegraphics[width=0.425\textwidth]{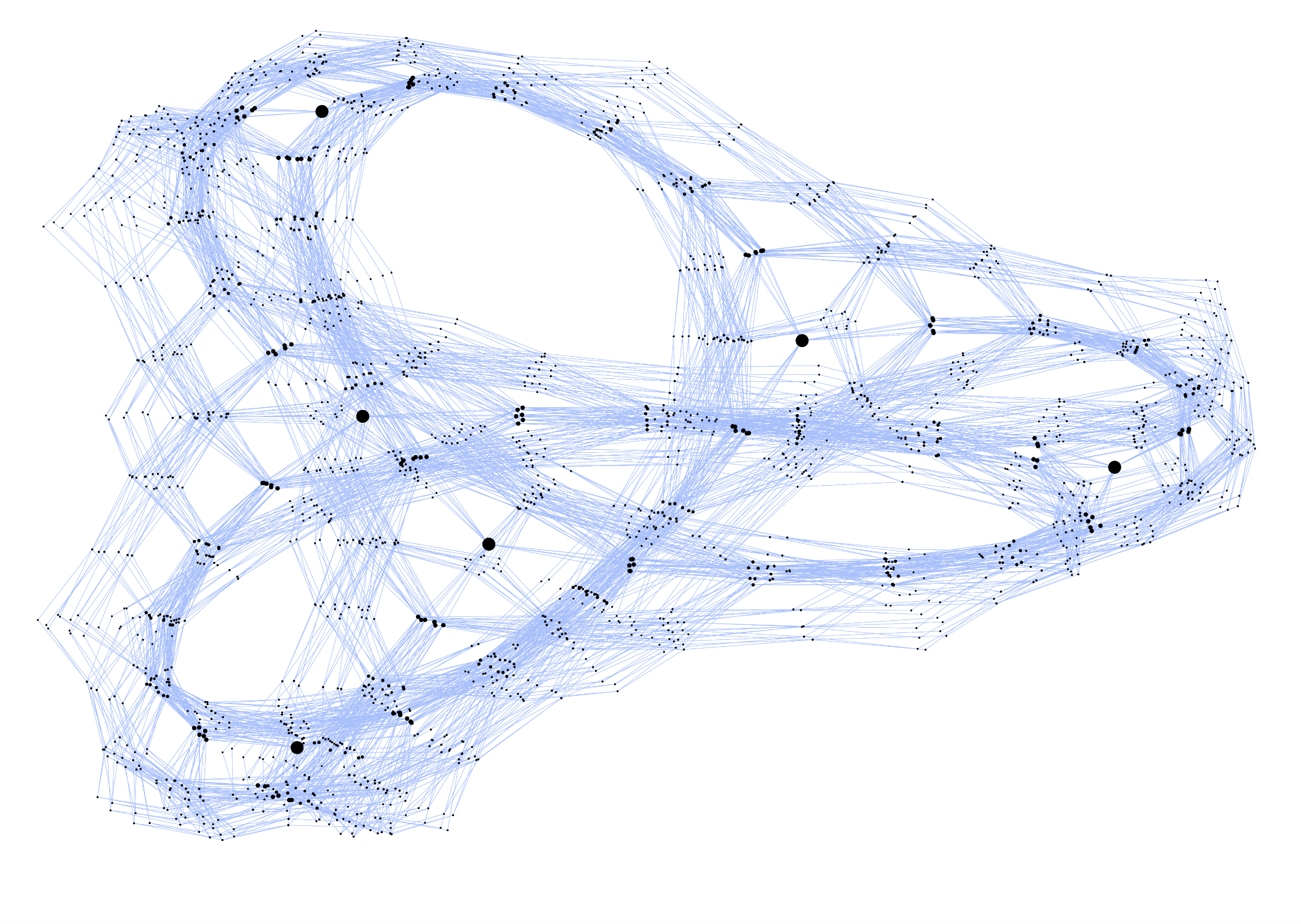}\label{fig:6x3_graphs-b}}
    \caption{(a)~Full coplanar ground-state phase-space network and (b)~short-loop network for a $6\times3$ kagome lattice ($N=54$ spins). Each node represents a coplanar ground-state configuration, and each edge represents a weathervane-loop rotation. The full network in (a) contains all 4752 coplanar configurations obtained by exact enumeration and forms a single Kempe sector for this finite lattice. The color of an edge denotes the corresponding barrier energy per spin $E_{\rm b}$, defined as the maximum zero-point energy encountered along the continuous weathervane-loop rotation path. (b)~The component reachable from the same $\sqrt{3}\times\sqrt{3}$ reference configuration using only elementary six-spin weathervane loops; all edges therefore correspond to the same lowest barrier-energy scale and are shown uniformly. The size of a node is indicative of its degree.}
    \label{fig:6x3_graphs}
\end{figure*}

Figure~\ref{fig:6x3_graphs} shows phase-space networks for a $6\times 3$ kagome lattice with periodic boundary conditions, where the edge colors represent the barrier energy $E_{\rm b}$ associated with the corresponding weathervane-loop rotation. For this finite lattice, exact enumeration gives 4752 coplanar three-coloring configurations. The full network (Fig.~\ref{fig:6x3_graphs-a}) is generated by starting from a $\sqrt{3}\times\sqrt{3}$ configuration and successively applying all allowed weathervane-loop moves. It contains all 4752 coplanar configurations, connected by 20 952 edges, and forms a single connected cluster. Thus, for this particular $6\times3$ lattice, all coplanar configurations belong to a single Kempe sector under arbitrary weathervane-loop moves. The full network exhibits pronounced heterogeneity, with two major interconnected basins separated by edges with relatively large barrier energies. A large, highly connected basin dominates the network, while a smaller basin is linked to it through low-degree bottleneck nodes that mediate transitions between regions of phase space and are associated with higher-energy weathervane loops.

In contrast, Fig.~\ref{fig:6x3_graphs-b} shows the short-loop network generated from the same $\sqrt{3}\times\sqrt{3}$ reference configuration when only elementary six-spin weathervane rotations are allowed. This component contains 2148 nodes and 6264 edges, reflecting the restriction to the lowest-energy dynamical processes. All edges in this restricted network correspond to six-spin weathervane rotations and therefore carry the same barrier-energy scale. The short-loop network includes only those coplanar states that are mutually reachable through sequences of such elementary loops. For example, $q=0$ states contain only system-spanning winding loops, rather than elementary six-spin loops, and are therefore absent from the short-loop network shown here. The reduced number of nodes thus reflects restricted dynamical accessibility rather than a reduction of the underlying ground-state manifold. Dynamics generated solely by elementary loops is therefore not ergodic within the coplanar manifold, even though the full $6\times3$ network forms a single Kempe sector when all weathervane loops are included. Longer weathervane loops reconnect states that are disconnected under elementary-loop dynamics. From this perspective, restricting the dynamics to elementary loops partitions the Kempe sector into short-loop dynamical components, while the full network restores connectivity by including longer collective rearrangements.

The two panels in Fig.~\ref{fig:6x3_graphs} are drawn using independent graph layouts, so node positions should not be compared directly between Figs.~\ref{fig:6x3_graphs-a} and \ref{fig:6x3_graphs-b}. Although our analysis in Fig.~\ref{fig:6x3_graphs-b} focuses on the representative short-loop network generated from a $\sqrt{3}\times\sqrt{3}$ reference configuration, the qualitative features identified here are expected to be generic consequences of restricting the dynamics to the lowest-energy collective modes. In the following sections, we show that statistical measures of connectivity, spectra, and scaling for larger system sizes both generalize and sharpen these observations.

\section{Phase-Space Connectivity}
\label{sec:connectivity}

A basic measure of connectivity in a phase-space network is the degree of a node, defined as the number of distinct ground-state configurations that can be reached from a given state by a single weathervane loop rotation~\cite{costa2007characterization}. Physically, the degree quantifies the number of dynamical pathways available from a configuration and thus provides a direct measure of its local mobility in phase space. States with high degree are connected to many nearby configurations and are dynamically flexible, whereas low-degree states are comparatively constrained and can act as bottlenecks for transitions between regions of phase space. The distribution of degrees across the network therefore encodes how accessibility and dynamical heterogeneity are organized within the coplanar ground-state manifold, and serves as a natural starting point for characterizing both local structure and large-scale connectivity~\cite{strogatz2001exploring,albert2002statistical}.

\begin{figure}[tbp!]
    \centering
    \subfloat[]{\includegraphics[width=0.45\textwidth]{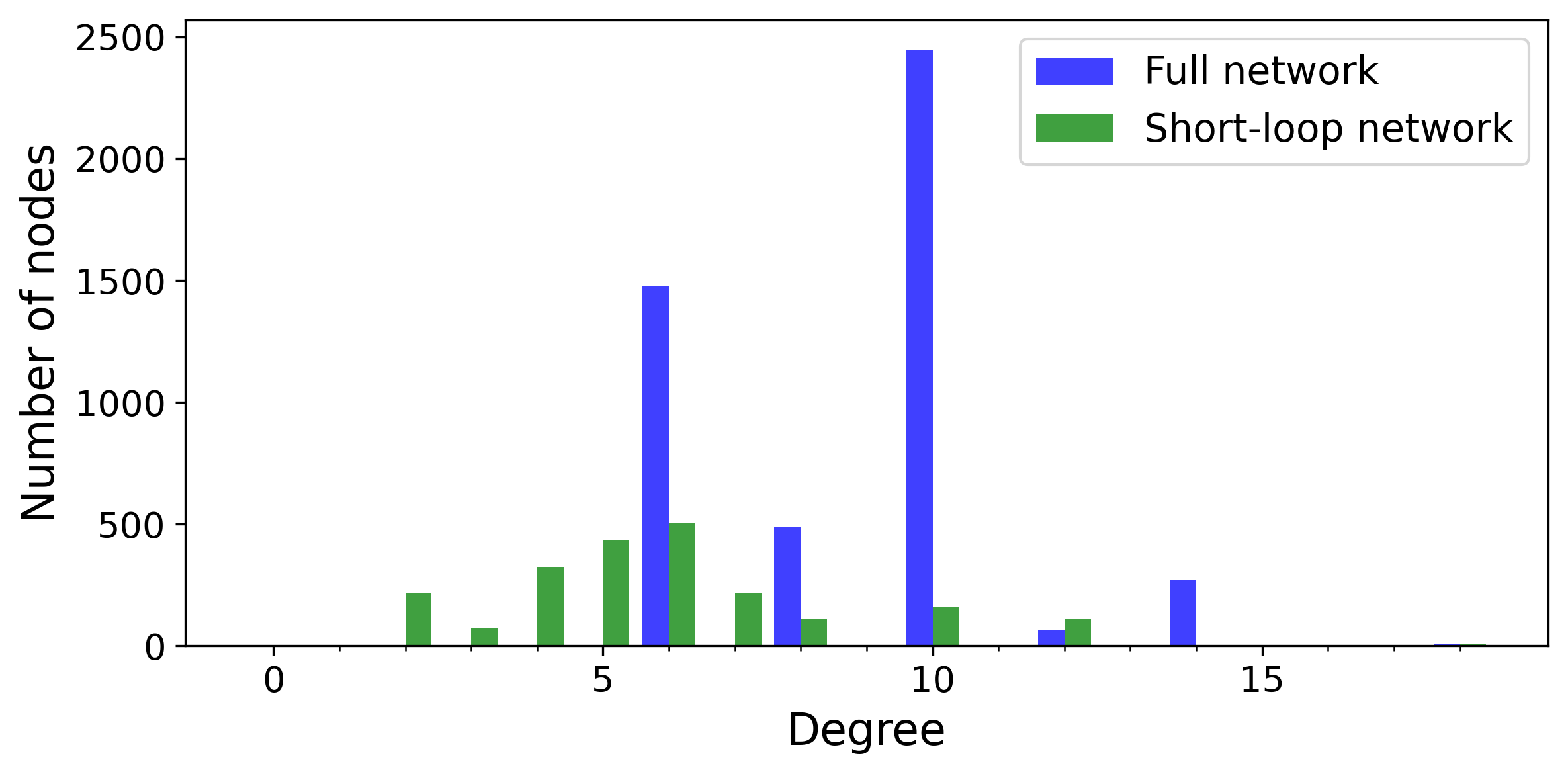}\label{fig:P(k)_exact-a}} \\
    \subfloat[]{\includegraphics[width=0.45\textwidth]{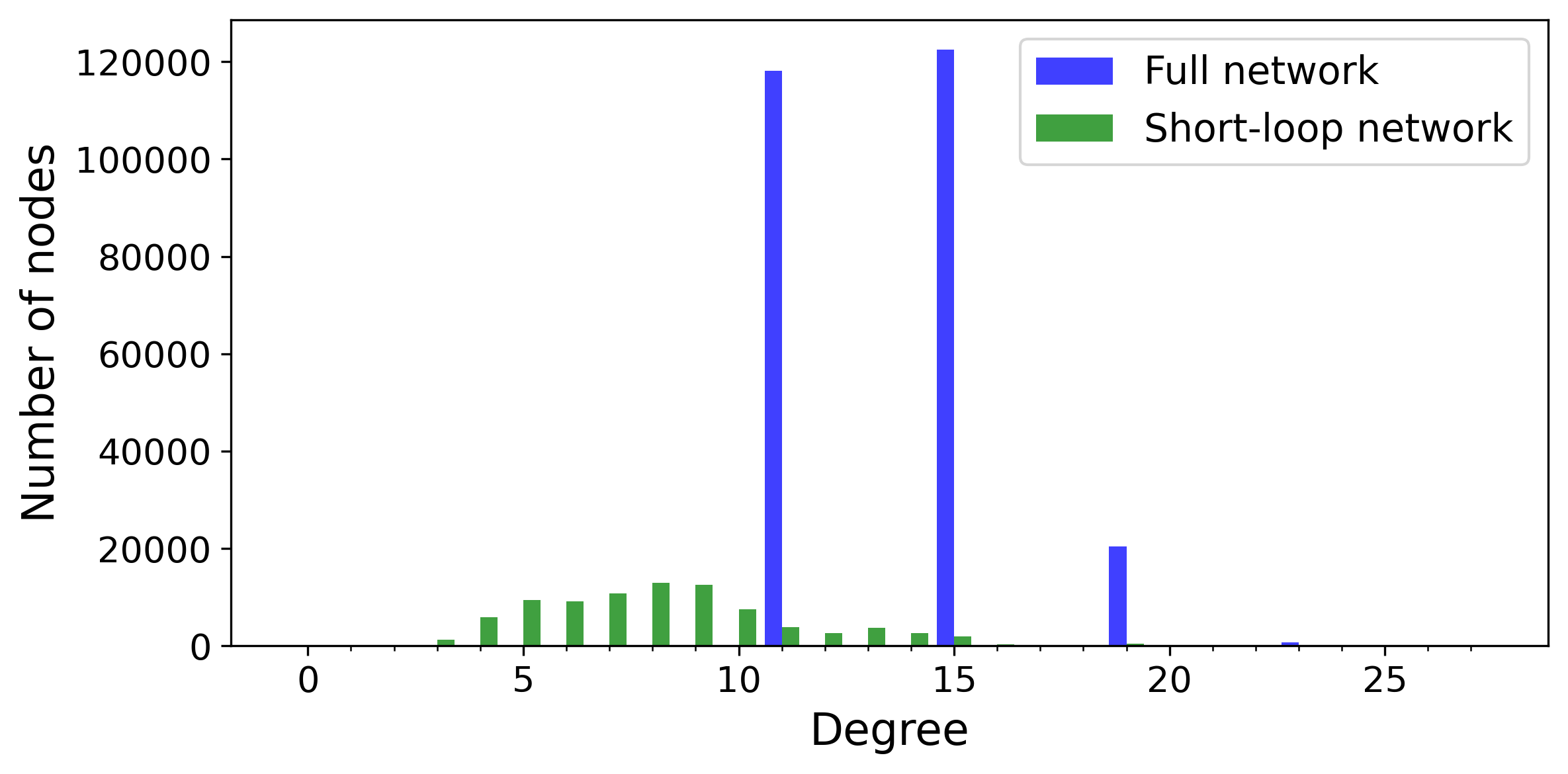}\label{fig:P(k)_exact-b}}
    \caption{Exact connectivity distributions for (a)~a $6\times 3$ kagome lattice ($N=54$ spins) and (b)~a $9\times 3$ kagome lattice ($N=81$ spins). Bars show the number of phase-space network nodes with each integer degree $k$ for the full network (blue) and the short-loop network (green). Degree values without bars have exactly zero node count. The blue and green bars are slightly displaced horizontally for visual clarity.}
    \label{fig:P(k)_exact}
\end{figure}

\subsection{Exact connectivity distributions}

For small system sizes, the phase-space networks can be constructed exactly by enumerating all coplanar ground-state configurations and identifying all allowed weathervane loop moves between them. From these networks, the exact connectivity (degree) distribution is obtained by counting, for each degree value $k$, the number of nodes that possess exactly $k$ distinct single-loop connections. Figure~\ref{fig:P(k)_exact} shows these distributions for a $6\times 3$ and $9\times 3$ kagome lattice with periodic boundary conditions, comparing the full network with the short-loop network. The distributions are plotted as histograms over integer degree values, so missing bars correspond to degree values with exactly zero node count. In particular, the sparse set of degree values visible in the full-network data reflects the restricted set of distinct loop counts allowed by the small finite geometries, rather than statistical noise or incomplete sampling.

As expected, the average degree in the full network is larger than in the short-loop network, reflecting the additional connectivity provided by longer weathervane loops. In these small systems, the two distributions also differ qualitatively in their support: the short-loop network exhibits nodes across a broader and more continuous set of degree values, whereas the full network has nodes concentrated at more widely spaced degree values. This spacing is a finite-size effect arising from the strong geometric constraints on long weathervane loops in small lattices, where the addition or removal of a loop changes the degree in relatively large discrete steps. Despite these differences, both networks contain exactly six states with degree $3L_xL_y$, corresponding to the $\sqrt{3}\times\sqrt{3}$ coplanar states. In these configurations, every hexagon forms a valid weathervane loop, making them maximally connected hubs of the phase-space network.

\begin{figure*}[htbp!]
    \centering
    \subfloat[]{\includegraphics[width=0.475\textwidth]{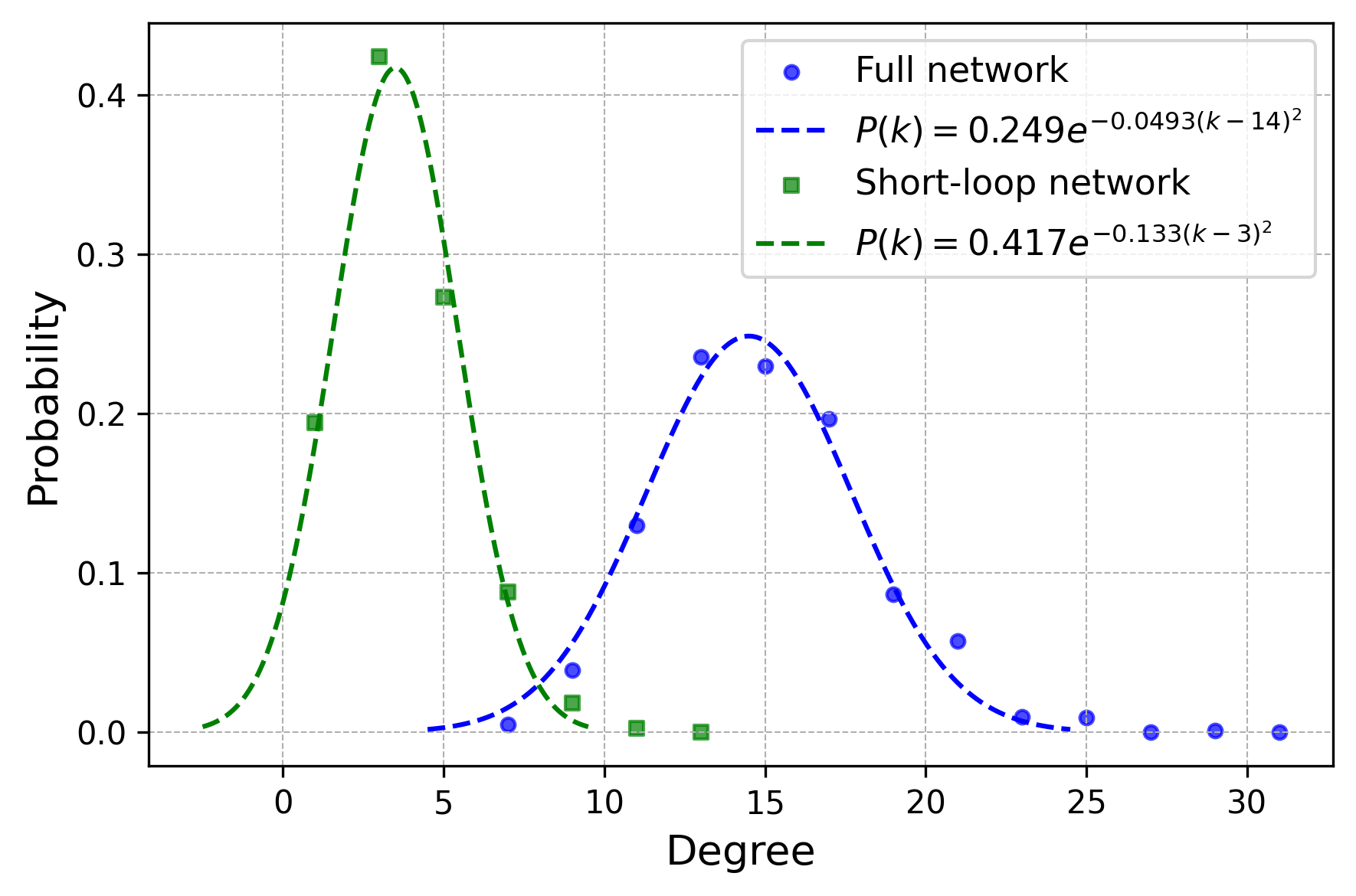}\label{fig:P(k)_dists-a}}
    \hfill
    \subfloat[]{\includegraphics[width=0.475\textwidth]{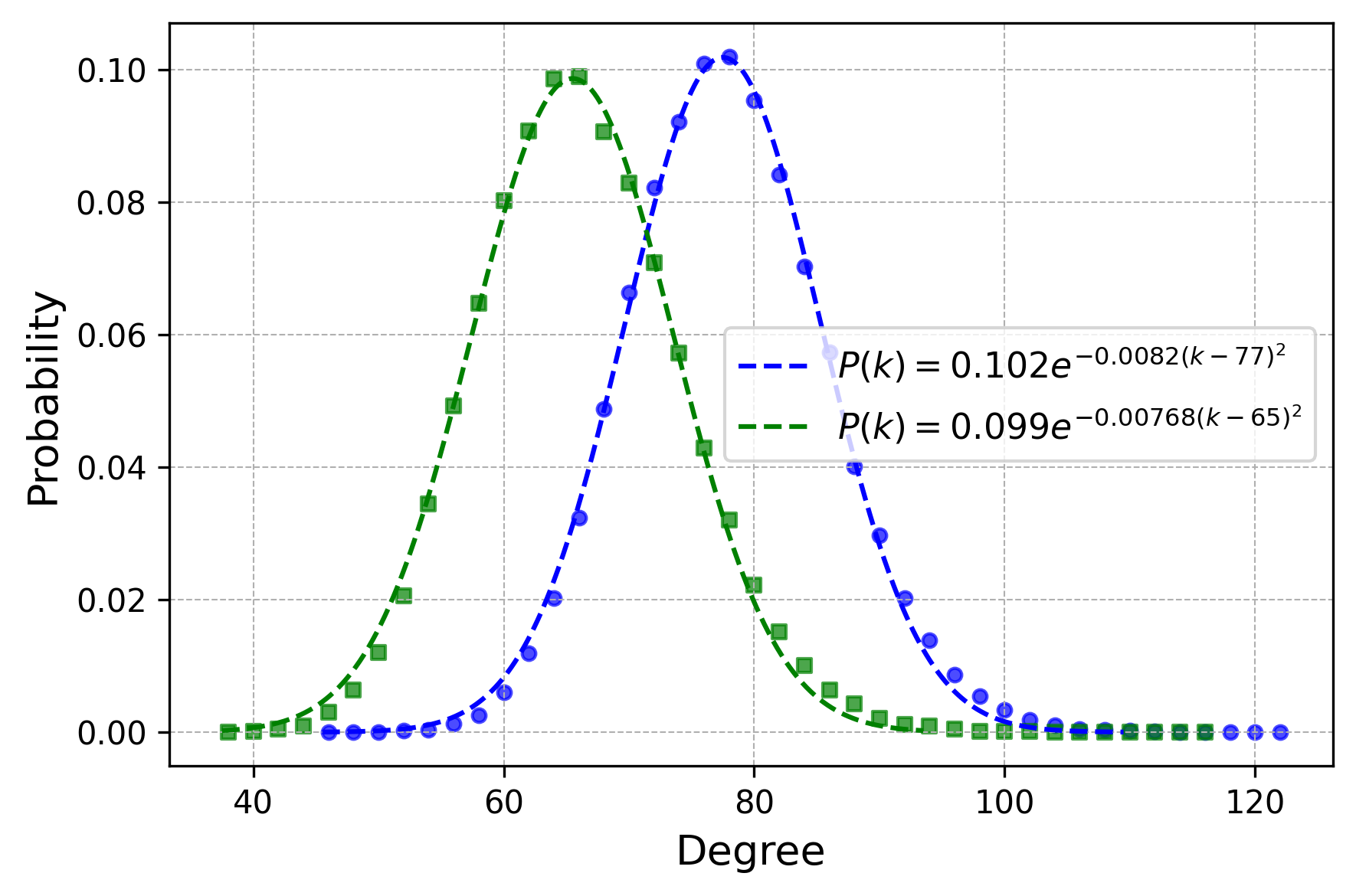}\label{fig:P(k)_dists-b}} \\
    \subfloat[]{\includegraphics[width=0.475\textwidth]{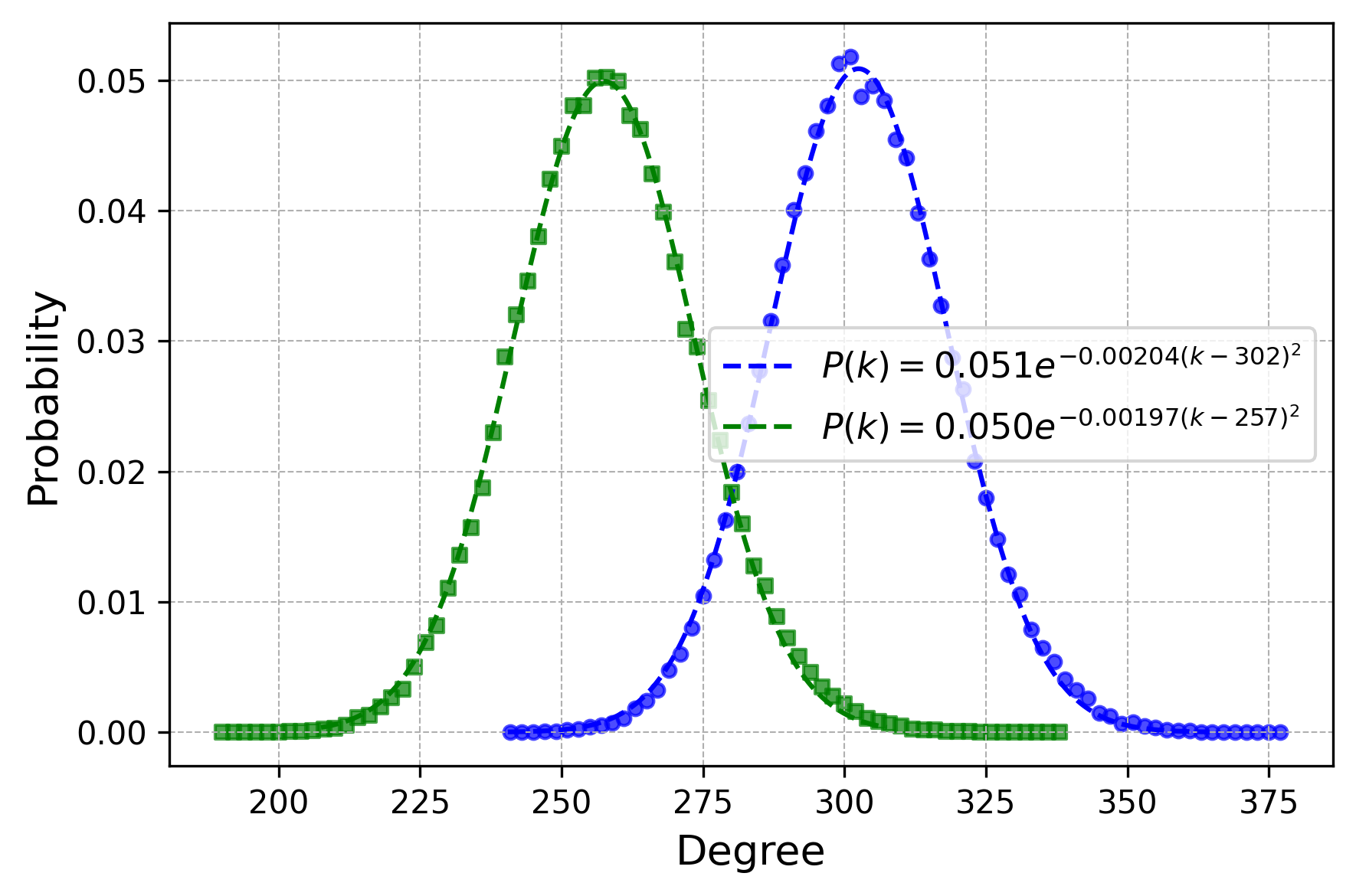}\label{fig:P(k)_dists-c}}
    \hfill
    \subfloat[]{\includegraphics[width=0.475\textwidth]{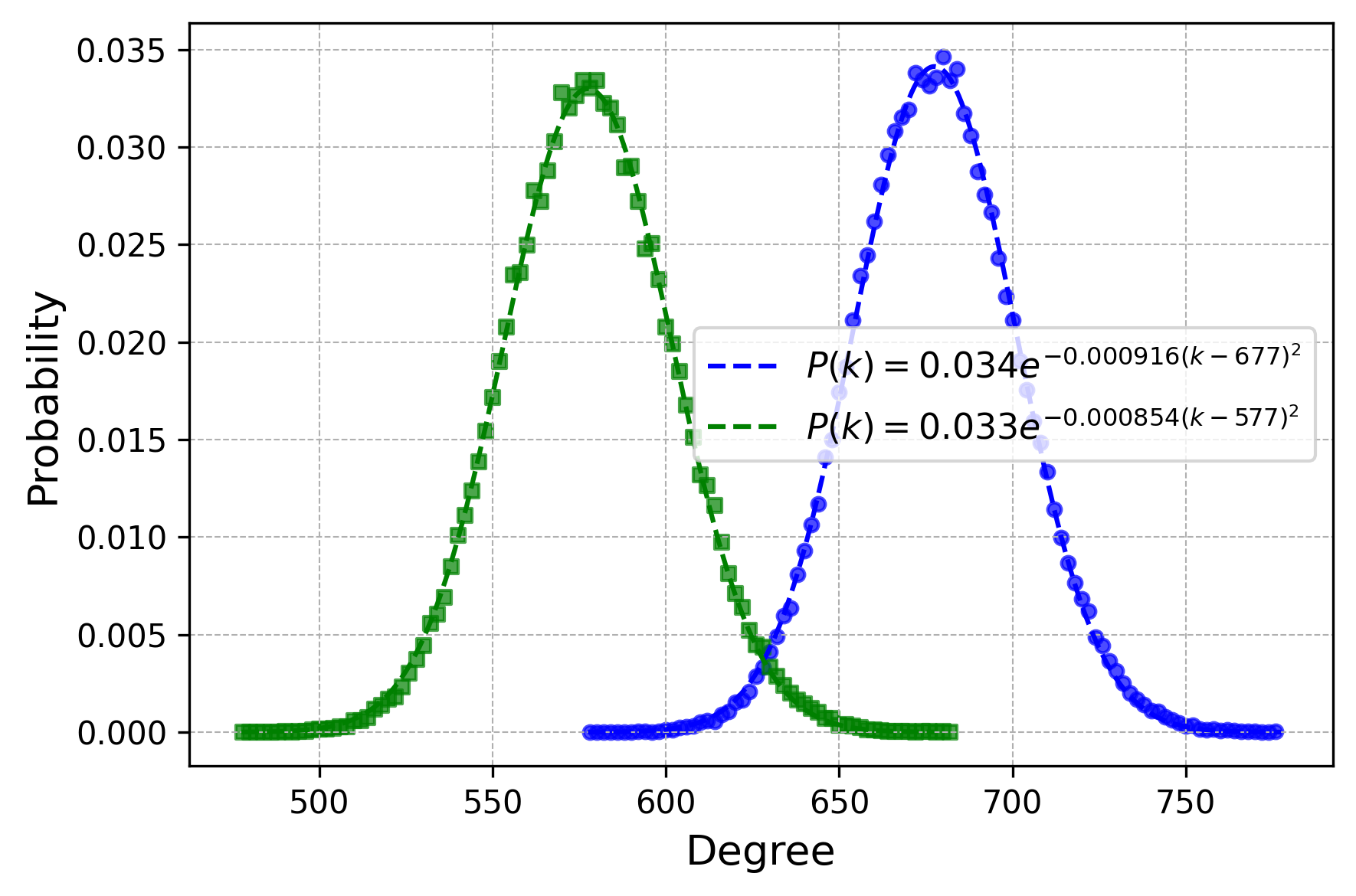}\label{fig:P(k)_dists-d}}
    \caption{Connectivity distributions for the full (blue) and short-loop (green) networks of (a)~$6\times 6$, (b)~$15\times 15$, (c)~$30\times 30$, and (d)~$45\times 45$ kagome lattices. Each data point represents the probability $P(k)$ of choosing a random configuration from the phase-space network that contains $k$ flippable weathervane loops. The data is fit with a Gaussian curve.}
    \label{fig:P(k)_distributions}
\end{figure*}

\subsection{Connectivity distributions of large networks}

We now shift from exact enumeration to a statistical characterization of phase-space connectivity. Rather than tracking individual nodes, we study the connectivity distribution $P(k)$, defined as the probability that a randomly chosen coplanar ground state admits exactly $k$ flippable weathervane loops.  The distribution $P(k)$ provides a compact description of connectivity that is well defined for larger systems, where exact enumeration is infeasible. This approach allows us to identify generic features of the coplanar manifold and distinguish finite-size effects from trends that persist as system size increases.

To estimate $P(k)$, we sample an ensemble of coplanar ground-state configurations $\{C_i\}$ generated by starting from a known coplanar state and performing a large number of random weathervane loop flips. For each configuration $C_i$, the number of flippable loops $k_i$ is determined by systematically searching for alternating loops on the lattice. The connectivity distribution is then obtained as a normalized histogram of the measured loop counts 
\begin{equation}
    P(k) = \frac{1}{N_{\mathrm{configs}}} \sum_{i=1}^{N_{\mathrm{configs}}} \delta_{k,k_i},
\end{equation}
which directly quantifies the likelihood that a randomly selected coplanar ground state has degree $k$.

\begin{figure*}[btp!]
    \centering
    \subfloat[]{\includegraphics[width=0.475\textwidth]{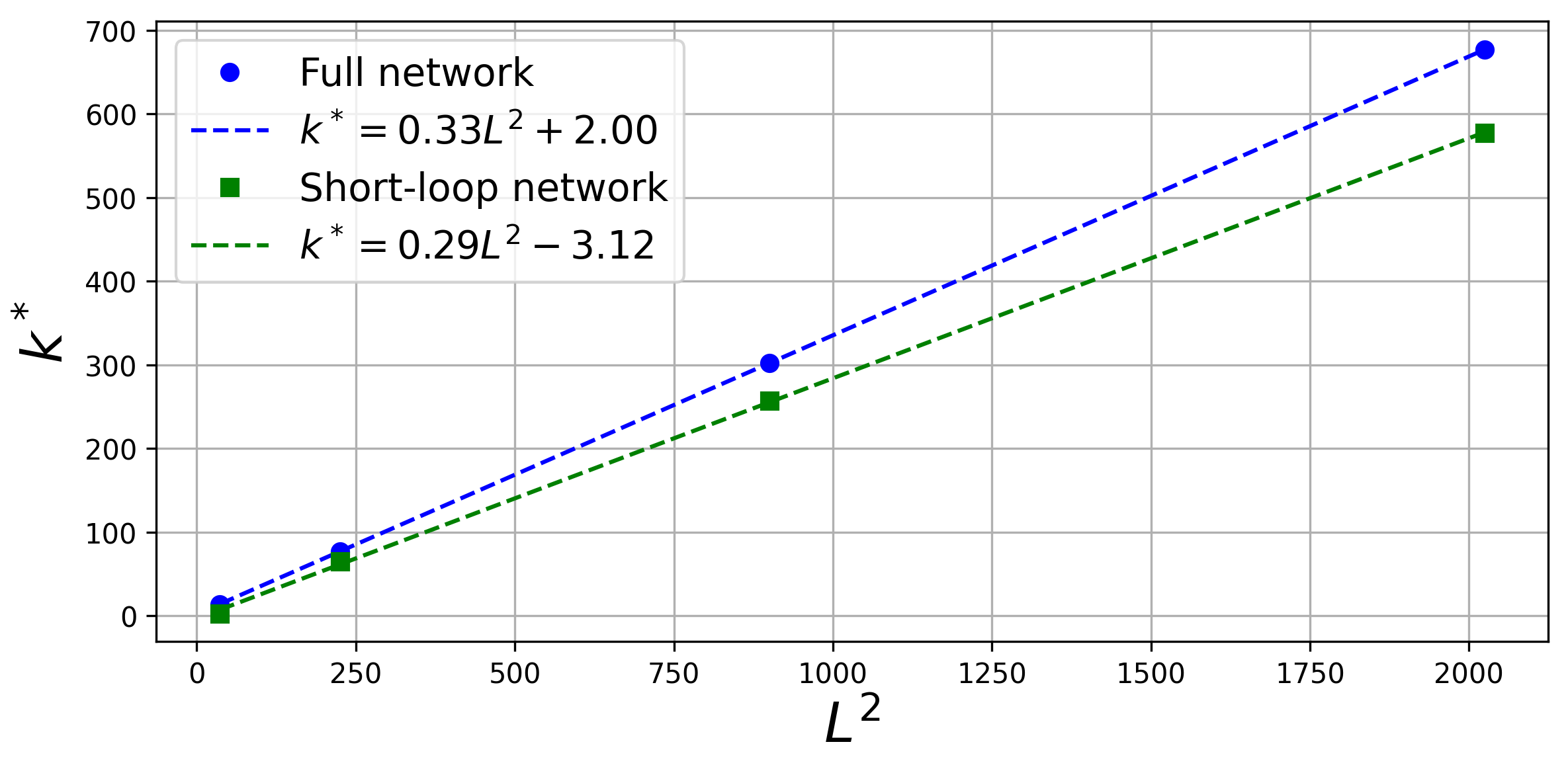}\label{kstar}}
    \hfill
    \subfloat[]{\includegraphics[width=0.475\textwidth]{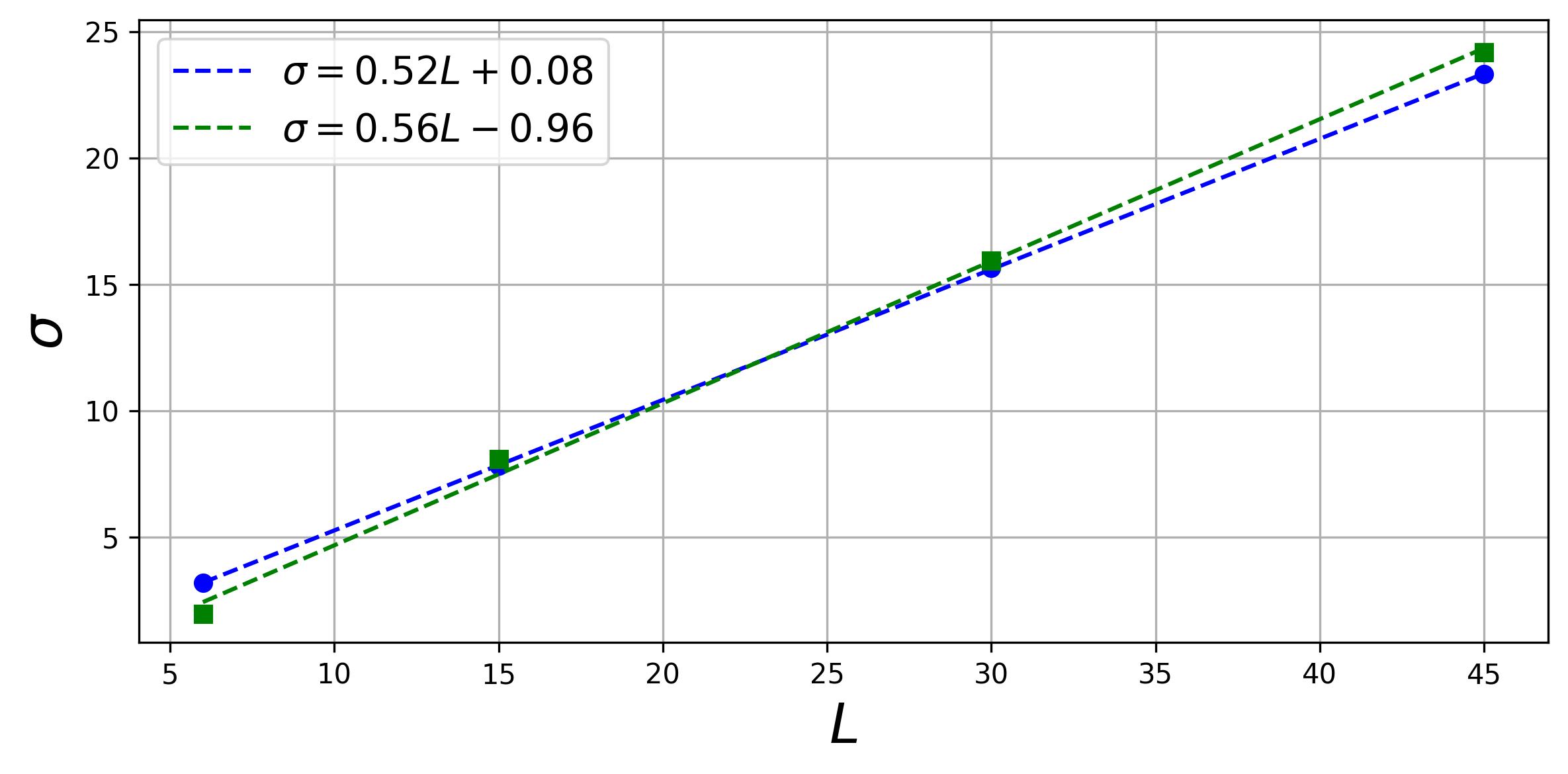}\label{sigma}}
    \caption{Fits for the scaling of (a)~the location of the $P(k)$ peak ($k^*$) and (b)~the width of the $P(k)$ Gaussian fit ($\sigma$) with the linear system size $L$. The fits show that the peak position scales linearly with the number of sites ($k^*\sim L^2$), the width scales linearly with the edge length ($\sigma\sim L$).}
    \label{fig:P(k)_stuff_fits}
\end{figure*}

Figure~\ref{fig:P(k)_distributions} shows the connectivity distributions $P(k)$ for both the full and short-loop networks across system sizes ranging from $6\times6$ to $45\times45$. The connectivity is bounded by two ordered limits: the $\sqrt{3}\times\sqrt{3}$ state, which maximizes the number of hexagonal loops and has the largest connectivity $k{\rm max}=3L^{2}$, and the $q=0$ state, which minimizes connectivity with $k_{\rm min}=3L$ system-spanning winding loops. For all system sizes, $P(k)$ is peaked at an intermediate connectivity and is strongly suppressed at both small and large $k$. For the finite sizes considered here, these two limits are not sharply separated, since the scales $3L$ and $3L^{2}$ are not yet widely separated at small~$L$.

Consistent with earlier studies~\cite{han2009phase,han2010phase}, the connectivity distribution $P(k)$ is well described by a Gaussian form for both networks. Gaussian fits reveal that the peak position scales extensively, $k^*\sim L^{2}$, while the width grows as $\sigma\sim L$ (Fig.~\ref{fig:P(k)_stuff_fits}). As a result, relative fluctuations vanish in the thermodynamic limit, $\sigma/k^* \sim 1/L$. This implies that almost all coplanar ground states possess a similar number of flippable loops, while highly ordered configurations such as the $q=0$ and $\sqrt{3}\times\sqrt{3}$ states form a vanishingly small fraction of the manifold.

The emergence of a Gaussian degree distribution has important implications for the structure of the coplanar manifold. It indicates the absence of hubs or heavy-tailed connectivity and suggests that, at the level of degree statistics, the phase-space network is effectively homogeneous~\cite{boccaletti2006complex}. Connectivity fluctuations arise from the accumulation of many weakly correlated local contributions, consistent with a central-limit-type mechanism~\cite{javarone2015gaussian}. As a result, dynamical heterogeneity is not driven by rare, highly connected states, but instead by more subtle correlations and constraints encoded in the network’s topology beyond degree alone.

Comparing the two networks reveals systematic differences. The short-loop network has a smaller mean connectivity $k^*$, reflecting the restriction to elementary six-spin loops, and its distribution is consistently shifted to lower $k$ relative to the full network. The increasing absolute separation between the full- and short-loop peak connectivities follows from their extensive scaling. Since the degree counts the number of available single-loop moves from a configuration, both $k^*_{\rm full}$ and $k^*_{\rm short}$ scale with the system area $L^2$. Writing $k^*_{\rm full}\simeq \rho_{\rm full}L^2$ and $k^*_{\rm short}\simeq \rho_6L^2$, where $\rho_6$ is the density of elementary six-spin loops and $\rho_{\rm full}-\rho_6$ is the density of additional longer-loop moves, we can see that their difference therefore also grows extensively: $k^*_{\rm full}-k^*_{\rm short}\simeq(\rho_{\rm full}-\rho_6)L^2$. Physically, this means that restricting to elementary loops removes an extensive set of possible rearrangements from the dynamics.

The width $\sigma$ of the short-loop distribution is also slightly larger relative to $L$, indicating enhanced variability in the number of flippable loops across configurations. This broader relative spread arises because short-loop connectivity is dominated by local hexagons, whose presence or absence produces stronger fluctuations, whereas the inclusion of longer loops in the full network smooths connectivity variations across the lattice. Consequently, the full network appears slightly more homogeneous, while the short-loop network retains greater configuration-to-configuration variability around its lower mean degree.

\section{Spectral Densities}
\label{sec:spectral}

The spectral properties of the coplanar ground-state network provide a global characterization of network structure that goes beyond local measures such as degree distributions~\cite{metz2020spectral}. While the degree captures immediate connectivity, the spectrum encodes how connectivity propagates through the network at all length scales.  Let $A$ denote the adjacency matrix of the network, with entries $A_{ij} = 1$ if configurations $C_i$ and $C_j$ are connected by a single weathervane loop rotation, and $A_{ij} = 0$ otherwise. The eigenvalues $\{\lambda_\alpha\}$ and associated eigenvectors $\ket{\psi_\alpha}$ of $A$ encode the collective connectivity structure: eigenvectors correspond to network eigenmodes, while eigenvalues measure how strongly these modes propagate across the network. The spectral density $\rho(\lambda)$ is defined as 
\begin{equation}
    \rho(\lambda) = \frac{1}{N}\sum_{\alpha = 1}^N\delta(\lambda-\lambda_\alpha)
\end{equation}
and provides a statistical characterization of the global connectivity of the network~\cite{farkas2001spectra,de2005spectral,metz2020spectral}.

A direct connection between the spectral density and network topology arises via closed walks. A closed walk of length $l$ is a sequence of $l$ successive weathervane loop flips that returns to the initial configuration $C_i$. Specifically, denoting $F_{\beta}$ as the operator corresponding to flipping loop $\beta$ (acting nontrivially only on configurations where $\beta$ is flippable), a closed walk $\mathcal{W} = \{\beta_1,\beta_2,\hdots,\beta_l\}$ satisfies
\begin{equation}
    F_{\beta_l}\cdots F_{\beta_2}F_{\beta_1}\ket{C_i} = \ket{C_i}.
\end{equation}
The trace of powers of the adjacency matrix counts these walks:
\begin{equation}
    \Tr(A^l) = \sum_\alpha\lambda_\alpha^l,
\end{equation}
so the $l$th moment $\mu_l = N^{-1}\Tr(A^l)$ of $\rho(\lambda)$ enumerates closed walks of length $l$ in the network. Each walk corresponds to a specific local network motif, such as an elementary hexagon or a combination of loops, linking the spectral moments to the network’s local connectivity structure. Short motifs dominate low-order moments, while longer or composite motifs contribute to higher-order moments and shape the tails of the spectral distribution.

Physically, the spectrum reflects how the network mediates transitions through phase space. Large eigenvalues correspond to highly connected collective modes $\ket{\psi_\alpha}$ that traverse many configurations, whereas small eigenvalues correspond to localized or weakly connected modes. The sign of $\lambda_\alpha$ distinguishes between correlated and anti-correlated patterns of connectivity. Therefore, the shape of $\rho(\lambda)$ encodes the topological organization of the network: local connectivity through individual weathervane flips and global features such as highly connected hubs or sparsely connected regions.

\begin{figure}[tbp!]
    \centering
    \subfloat[]{\includegraphics[width=0.445\textwidth]{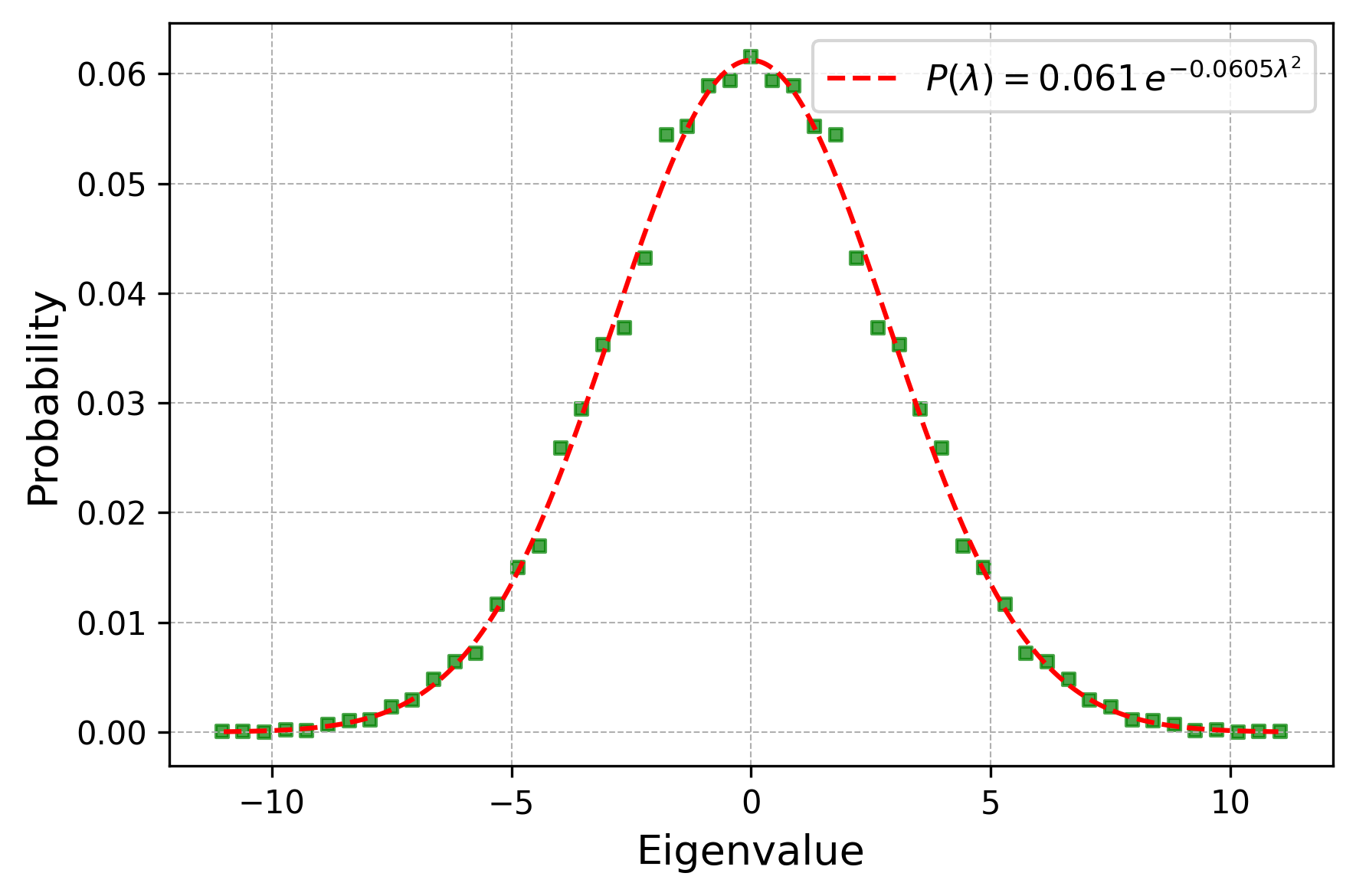}\label{spectral-a}} \\
    \subfloat[]{\includegraphics[width=0.445\textwidth]{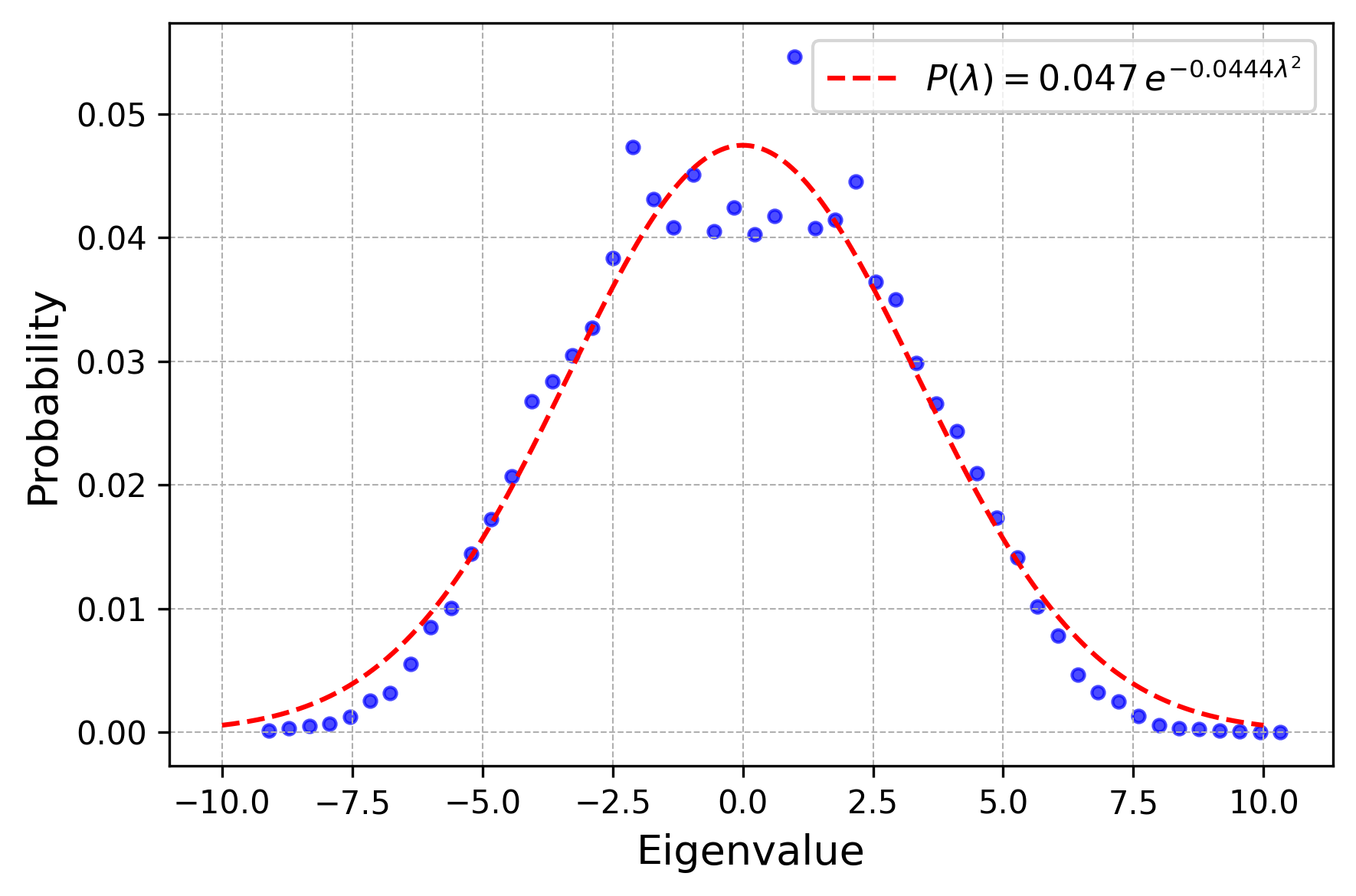}\label{spectral-b}}
    \caption{Spectral densities for representative phase-space networks of comparable graph size. (a) Short-loop network for a $9\times 3$ kagome lattice with periodic boundary conditions, containing 85 164 nodes. (b) Full network for a $4\times7$ kagome lattice with twisted boundary conditions, containing 72 960 nodes. The short-loop spectrum is close to Gaussian, while the full-network spectrum shows non-Gaussian features.}
    \label{fig:spectral_density}
\end{figure}

Figure~\ref{fig:spectral_density} compares representative spectral densities of the short-loop and full phase-space networks. Since the spectral density is obtained from the adjacency matrix, the smoothness and resolution of $\rho(\lambda)$ depend directly on the number of nodes in the phase-space network. We therefore choose examples whose phase-space networks have comparable graph sizes. The comparison in Fig.~\ref{fig:spectral_density} should thus be understood as a comparison between representative short-loop and full networks of similar adjacency-matrix dimension, rather than a same-lattice finite-size comparison.

Figure~\ref{spectral-a} shows the short-loop network for a $9\times3$ kagome lattice with periodic boundary conditions, comprising 85 164 nodes. As in other systems where dynamics is restricted to the smallest loop updates~\cite{han2009phase,han2010phase}, the spectral density closely follows a Gaussian form. This behavior arises from the effective statistical independence of local weathervane flips in large networks: in the thermodynamic limit, the dominant contributions to the spectral moments come from many weakly correlated sequences of elementary loop flips. As a result, all moments of the spectral distribution converge to those of a Gaussian, with vanishing odd moments and even moments determined solely by the variance.

Even for the finite system shown in Fig.~\ref{spectral-a}, the spectral density is already nearly Gaussian, indicating a high degree of statistical homogeneity in the short-loop network. Such a Gaussian spectrum is characteristic of networks in which connectivity fluctuations are self-averaging and no rare, highly connected structures dominate. Dynamically, this implies that random walks or diffusion on the network are governed by typical pathways rather than bottlenecks, leading to normal diffusive behavior in phase space with a single dominant relaxation scale, rather than anomalous or hierarchical dynamics.

The spectral density of a representative full network is shown in Fig.~\ref{spectral-b}. This full network is constructed for a $4\times7$ kagome lattice with twisted boundary conditions and comprises 72 960 nodes, making it comparable in graph size to the short-loop network in Fig.~\ref{spectral-a}. The twisted boundary condition is used because the additional freedom in closing the finite lattice allows us to access a coloring sector whose phase-space graph size is comparable to that of the short-loop example. In particular, the twist relaxes some of the finite-size matching constraints imposed by ordinary periodic boundary conditions and makes it easier to accommodate ordered coloring patterns, such as $\sqrt{3}\times\sqrt{3}$ states. While the bulk of the spectral density remains approximately Gaussian, the weight near the edges of $\rho(\lambda)$ is noticeably suppressed. This behavior can be understood from the structure of the adjacency operator $A=\sum_\beta F_\beta$ and the statistics of closed walks it generates.

Large-magnitude eigenvalues $|\lambda_\alpha|$ correspond to eigenmodes $\ket{\psi_\alpha}$ that are coherently reinforced under repeated application of $A$, so that $A^l\ket{\psi_\alpha}=\lambda_\alpha^l\ket{\psi_\alpha}$ remains large for increasing $l$. Equivalently, these eigenvalues dominate the high-order moments $\mu_l=N^{-1}\Tr(A^l)$, which count closed walks of length $l$. In the short-loop network, where dynamics is generated solely by local, elementary moves, long closed walks decompose into many weakly correlated motifs. This allows contributions to $\Tr(A^l)$ to factorize, producing central-limit behavior and a Gaussian spectral density.

In contrast, closed walks in the full network involve products of loop-flip operators $F_{\beta_l}\cdots F_{\beta_1}$ drawn from a broad distribution of loop lengths. Short loops induce local rearrangements, while long loops act collectively over extended regions of the lattice, and the corresponding operators generally do not commute. As a result, successive applications of $F_\beta$ introduce scale-dependent sign and amplitude changes that cannot be coherently aligned across all loop lengths. Walks that mix short and long loops therefore fail to reinforce a single eigenmode: contributions that enhance a mode under short-loop updates are typically partially canceled by long-loop updates, and vice versa. This incompatibility suppresses the accumulation of large-$|\lambda_\alpha|$ contributions and reduces the growth of $\Tr(A^l)$ relative to the uncorrelated case.

From this perspective, the depletion of spectral weight near the edges of $\rho(\lambda)$ reflects the absence of eigenmodes that are simultaneously stable under the full hierarchy of loop updates. Modes adapted to short-loop constraints develop fine-scale structure and are highly sensitive to nonlocal long-loop flips, while modes smooth enough to survive long-loop updates are disrupted by proliferating short-loop rearrangements. Because loop-flip operators of different lengths do not commute, these constraints cannot be satisfied simultaneously. Consequently, only modes associated with intermediate scales retain partial coherence under repeated application of $A$, populating the bulk of the spectrum, while modes that would produce extremal eigenvalues are statistically suppressed.

\section{Fractal Properties}
\label{sec:fractal}

Beyond local connectivity and spectral properties, the structure of the coplanar ground-state manifold can be characterized through its fractal organization in phase space~\cite{gallos2007review,fronczak2024scaling,bunimovich2024fractal}. To this end, we estimate an effective box-counting dimension of the phase-space network, which quantifies how the number of boxes required to cover the network decreases as the graph-theoretic length scale is increased. Physically, this box-counting analysis characterizes the efficiency with which phase space is explored under successive weathervane loop flips, providing a global measure of connectivity that is sensitive to collective dynamical constraints rather than local degree statistics.

We employ the standard box-covering method adapted to networks, in which the distance between two nodes is defined as the length of the shortest path connecting them. For a given box size $l_B$, defined as the maximum graph distance allowed within a box, the network is covered by the minimal number of boxes $N_B(l_B)$. Many complex networks exhibit a power-law relation~\cite{song2005self},
\begin{equation}
N_B(l_B)\sim l_B^{-d_B},
\end{equation}
which signals self-similarity under coarse-graining and defines the fractal box dimension $d_B$~\cite{hutchinson1981fractals,theiler1990estimating}. In the finite networks studied below, we use this algebraic form as a finite-range fitting description and interpret the fitted exponent as an effective box-counting exponent. Since determining the optimal covering is computationally prohibitive for large networks, we estimate $N_B(l_B)$ using a randomized box-covering procedure in which boxes are grown around randomly selected nodes and iteratively placed until all nodes are covered. Averaging over many realizations yields a robust estimate of $N_B(l_B)$ and its scaling behavior~\cite{peng2011self}.

\begin{figure}[tbp!]
    \centering
    \includegraphics[width=\linewidth]{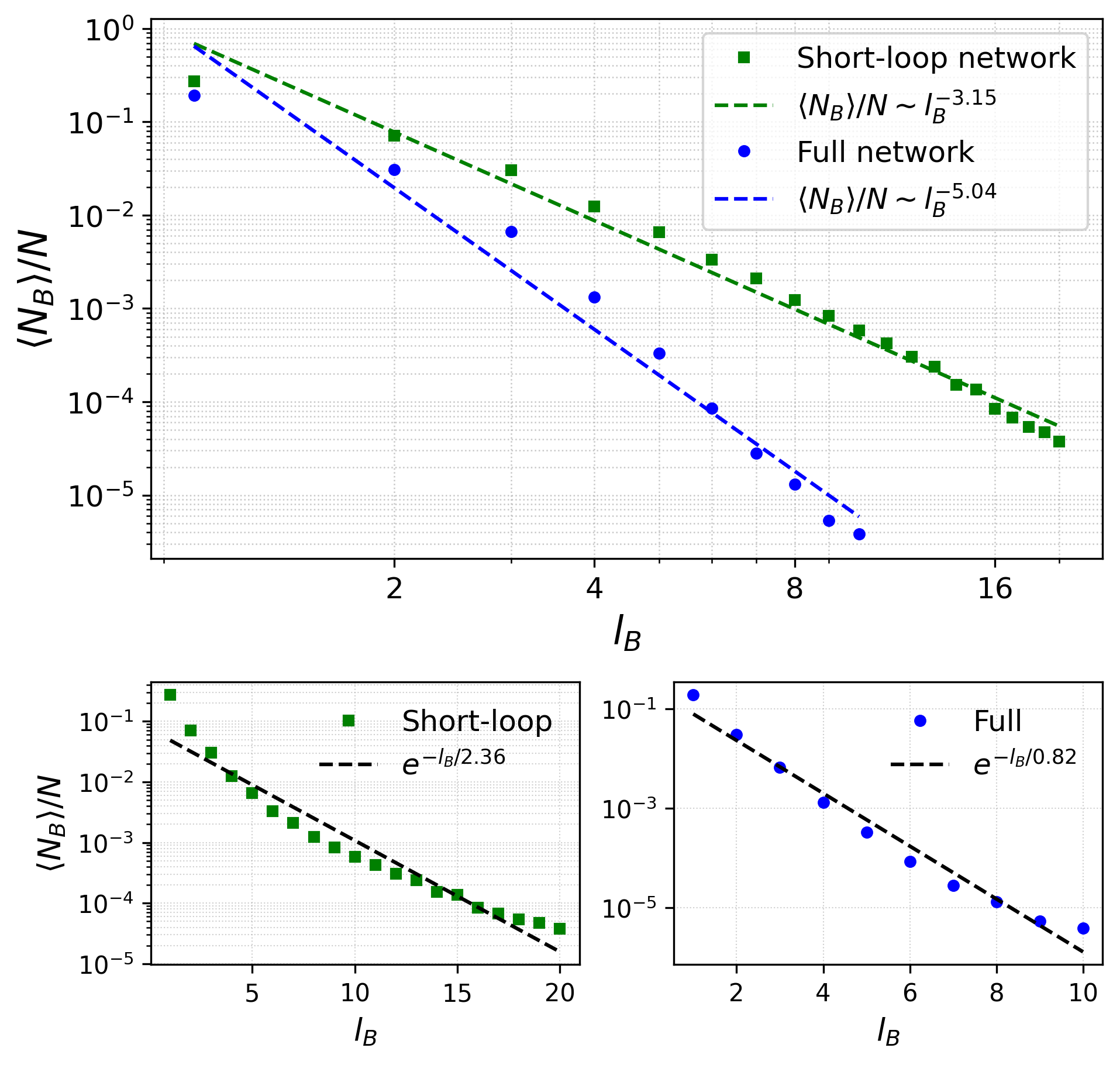}
    \caption{Fractal box-counting analysis of the full (blue) and short-loop (green) phase-space networks for a $9\times3$ kagome lattice. The top panel shows the normalized number of boxes $\langle N_B\rangle/N$ required to cover the network as a function of box size $l_B$ on log-log axes, where distances are measured in graph-theoretic units. Power-law fits are shown for comparison. The lower panels show the same data on semilogarithmic axes together with exponential fits. The fits are used to compare algebraic-like and exponential-like descriptions over the accessible range of box sizes, and quantitative fit metrics are reported in Table~\ref{tab:box_fit}.}
    \label{fig:fractal_dim}
\end{figure}

Figure~\ref{fig:fractal_dim} presents the box-counting analysis for both the short-loop and full phase-space networks of a $9\times3$ lattice with periodic boundary conditions. The top panel shows $\langle N_B\rangle/N$ as a function of $l_B$ on log-log axes, while the lower panels show the same data on semilogarithmic axes. This combination provides a visual comparison between algebraic and exponential fitting descriptions over the accessible finite range of box sizes. The exponential form is included as a comparison because networks with shortcuts or small-world-like connectivity can exhibit an approximately exponential decrease of $N_B(l_B)$ under box covering, rather than the algebraic behavior expected for a fractal network~\cite{rozenfeld2010small}.

Table~\ref{tab:box_fit} provides a quantitative comparison between the two finite-range fitting descriptions used in Fig.~\ref{fig:fractal_dim}. Because $\langle N_B\rangle/N$ spans several orders of magnitude, the fits are performed in logarithmic space. We define the log-space root-mean-square error as
\begin{equation}
    \mathrm{RMSE}_{\log} = \sqrt{\frac{1}{n}\sum_{i=1}^{n}\left[\log y_i - \log y_i^{\rm fit}\right]^2},
\end{equation}
where $y_i=\langle N_B(l_{B,i})\rangle/N$ and $n$ is the number of box sizes included in the fit. This metric measures the typical multiplicative deviation between the data and the fitted form. To compare the algebraic and exponential fitting forms in a standard model-selection framework, we also report the small-sample corrected Akaike information criterion, AICc~\cite{akaike1974new,hurvich1989regression,burnham2002model}. For a fit with residual sum of squares $\mathrm{RSS}$ in log space and $p$ fitted parameters, we use
\begin{equation}
    \mathrm{AICc} = n\log\left(\frac{\mathrm{RSS}}{n}\right) +2p +\frac{2p(p+1)}{n-p-1}.
\end{equation}
We report $\Delta\mathrm{AICc}=\mathrm{AICc}_{\rm exp}-\mathrm{AICc}_{\rm pow}$, so that positive values favor the algebraic fit and negative values favor the exponential fit. Since both forms have the same number of fitted parameters, this comparison mainly reflects their relative residual errors, with a correction for the limited number of available box sizes.

\begin{table}[tb]
    \centering
    \renewcommand{\arraystretch}{1.3}
    \caption{Fit-quality comparison for the box-counting data in Fig.~\ref{fig:fractal_dim}. Algebraic fits use $\langle N_B\rangle/N \sim l_B^{-d_B^{\rm eff}}$, while exponential fits use $\langle N_B\rangle/N \sim \exp(-l_B/\xi^{\rm eff})$. Fits are performed in log space, and RMSE$_{\rm pow}$ and RMSE$_{\rm exp}$ give the corresponding log-space root-mean-square errors. The quantity $\Delta\mathrm{AICc}=\mathrm{AICc}_{\rm exp}-\mathrm{AICc}_{\rm pow}$ compares the small-sample corrected Akaike information criterion for the two fits, where positive values favor the algebraic fit and negative values favor the exponential fit.
    \label{tab:box_fit}}
    \begin{ruledtabular}
    \begin{tabular}{lccccc}
        Network & $d_B^{\rm eff}$ & $\xi^{\rm eff}$ & RMSE$_{\rm pow}$ & RMSE$_{\rm exp}$ & $\Delta\mathrm{AICc}$ \\
        \hline
        Short-loop & $3.15$ & $2.36$ & $0.330$ & $0.628$ & $+25.7$ \\
        Full & $5.04$ & $0.82$ & $0.654$ & $0.588$ & $-2.1$ \\
    \end{tabular}
    \end{ruledtabular}
\end{table}

The quantitative fit comparison in Table~\ref{tab:box_fit} shows that, over the accessible range of $l_B$, the short-loop network is consistent with fractal-like finite-size behavior and is better captured by an algebraic form than by an exponential. The log-space error is substantially smaller for the algebraic fit, $\mathrm{RMSE}_{\rm pow}=0.330$, than for the exponential fit, $\mathrm{RMSE}_{\rm exp}=0.628$, and the AICc difference $\Delta\mathrm{AICc}=+25.7$ favors the algebraic description. We therefore interpret the fitted value $d_B^{\rm eff}\approx 3.15$ as an effective finite-size box-counting exponent for the short-loop network, rather than as a definitive asymptotic fractal dimension. This fractal-like behavior indicates that, although the degree distribution is approximately Gaussian and local connectivity is statistically homogeneous, the global organization of the short-loop network remains highly constrained. In particular, restricting the dynamics to elementary six-spin weathervane loops removes long-loop shortcuts in configuration space, so exploration proceeds through sequences of local rearrangements.

For the full network, the fit comparison is less decisive. The exponential fit gives a slightly smaller log-space error than the algebraic fit, with $\mathrm{RMSE}_{\rm exp}=0.588$ compared with $\mathrm{RMSE}_{\rm pow}=0.654$, and $\Delta\mathrm{AICc}=-2.1$. The full-network data therefore show only a weak finite-range preference for an exponential-like description, and we do not interpret them as evidence for a clean exponential box-counting law. At the same time, the algebraic fit gives a larger effective exponent, $d_B^{\rm eff}\approx 5.04$, than in the short-loop network. This larger finite-range exponent reflects the more rapid decrease of $\langle N_B\rangle/N$ with $l_B$ when longer weathervane loops are included. Physically, these longer loops connect configurations that would otherwise require many elementary loop flips, thereby acting as effective shortcuts in the phase-space network.

Taken together, although the box-counting results do not identify clear asymptotic scalings, they do indicate a relative geometric distinction between the two networks. The short-loop network exhibits more algebraic-like finite-size behavior, consistent with constrained and hierarchical exploration of configuration space. Including longer weathervane loops makes the full network more compact in graph distance, as reflected either by the weak preference for an exponential-like fit or, within an algebraic description, by the larger effective exponent $d_B^{\rm eff}$. This weakens the short-loop hierarchical structure and enables more rapid global connectivity across the coplanar manifold.

\section{Conclusions}
\label{sec:conclusions}

In this work, we constructed phase-space networks for the coplanar ground-state manifold of the kagome Heisenberg antiferromagnet, with nodes representing coplanar configurations and edges corresponding to weathervane-loop transitions. By comparing full networks, which include all allowed loop updates, to short-loop networks restricted to elementary six-spin rotations, we isolated how energetic constraints and collective dynamics reshape phase-space connectivity. Degree statistics show that in large systems the connectivity distribution $P(k)$ is well described by a Gaussian, with a mean scaling as $k^*\sim L^2$ and a width $\sigma\sim L$, implying that relative fluctuations in connectivity vanish in the thermodynamic limit. Restricting the dynamics to short loops lowers the typical connectivity and modestly enhances heterogeneity, reflecting the reduced set of available local rearrangements.

Global probes reveal that the hierarchy of loop updates has an even more pronounced impact on the large-scale organization of phase space. The short-loop network exhibits a Gaussian spectral density, consistent with closed-walk statistics dominated by weakly correlated local motifs and indicative of statistically homogeneous, diffusive dynamics. In contrast, the full network displays suppressed spectral weight at large $|\lambda|$, reflecting correlations induced by the coexistence of short and long weathervane loops and the incompatibility of local and nonlocal rearrangements. Box-counting analysis further shows a relative geometric distinction between the two networks. The short-loop network is better described by an algebraic, fractal-like finite-size form over the accessible range, whereas the full network shows a more rapid reduction of graph distances, with a weak finite-range preference for an exponential-like fit. This behavior is consistent with the emergence of effective shortcuts in configuration space and a crossover toward a more rapidly connected, \mbox{small-world-like} organization when longer loops are included.

Taken together, our results demonstrate that the coplanar ground-state manifold of the kagome antiferromagnet forms a highly constrained phase space whose structure is controlled by the hierarchy of available loop updates and their associated energy scales. Even in the absence of strong degree heterogeneity, restricting dynamics to elementary loops produces fractal-like finite-size box-counting behavior, reflecting constrained and hierarchical exploration. Including longer weathervane loops qualitatively alters this picture by introducing nonlocal connections that weaken the short-loop hierarchy and produce a more compact global organization of the phase-space
network. Although the present work is classical, the resulting phase-space network may also provide a useful semiclassical framework for organizing quantum tunneling or resonance processes between coplanar configurations in large-$S$ kagome antiferromagnets. For small $S$, where quantum fluctuations are stronger and the ground state need not be localized near individual classical coplanar configurations, this network should instead be viewed as a structural reference for the constrained configuration space rather than as a quantitative description of the quantum dynamics. More broadly, our phase-space network framework provides a unified way to connect microscopic dynamical constraints to emergent global structure, and offers a natural starting point for understanding slow dynamics, trapping, and ergodicity breaking in frustrated magnets and other constrained many-body systems.

\begin{acknowledgments}
This work was supported by the U.S. Department of Energy, Office of Science, Basic Energy Sciences, through Grant No. DE-SC0026087.
\end{acknowledgments}

\bibliography{refs}% Produces the bibliography via BibTeX.

\end{document}